\documentclass[prb,aps,showpacs,superscriptaddress,twocolumn]{revtex4}
\usepackage{color} 
\usepackage{times}
\usepackage{amssymb} 
\usepackage{amsmath} 
\usepackage{ragged2e}
\usepackage{graphicx}
\usepackage{epstopdf}
\usepackage[english]{babel}
\usepackage{mathrsfs}
\usepackage{textcomp}
\usepackage{amsthm}
\usepackage{amsfonts}
\usepackage{soul}
\usepackage{braket} 

\newcommand{\be}{\begin{equation}}
\newcommand{\ee}{\end{equation}}
\newcommand{\bea}{\begin{eqnarray}}
\newcommand{\eea}{\end{eqnarray}}
\newcommand{\ba}{\begin{array}}
\newcommand{\ea}{\end{array}}
\newcommand{\lf}{\lfloor}
\newcommand{\rf}{\rfloor}
\newcommand{\change}[1]{\textcolor{black}{#1}}
\newcommand{\changeb}[1]{\textcolor{black}{#1}}

\begin{document}

\title{Violation of cluster decomposition and absence of light-cones in 
local integer and half-integer spin chains}
\author{L. Dell'Anna}
\affiliation{Dipartimento di Fisica e Astronomia ``G. Galilei'', 
Universit\`a di Padova, via F. Marzolo 8, I-35131, Padova, Italy}
\author{O. Salberger}
\affiliation{C. N. Yang Institute for Theoretical Physics, 
Stony Brook University, NY 11794, USA}
\author{L. Barbiero}
\affiliation{Dipartimento di Fisica e Astronomia ``G. Galilei'', 
Universit\`a di Padova, via F. Marzolo 8, I-35131, Padova, Italy}
\affiliation{CNR-IOM DEMOCRITOS Simulation Center, Via Bonomea 265, I-34136
Trieste, Italy}
\author{A. Trombettoni}
\affiliation{CNR-IOM DEMOCRITOS Simulation Center, Via Bonomea 265, I-34136
Trieste, Italy}
\affiliation{SISSA and INFN, Sezione di Trieste, Via Bonomea 265, I-34136 
Trieste, Italy}
\author{V. E. Korepin}
\affiliation{C. N. Yang Institute for Theoretical Physics, 
Stony Brook University, NY 11794, USA}


\begin{abstract}
We compute the {\change {ground state}} correlation functions of 
an exactly solvable chain of integer spins, recently introduced in 
[R. Movassagh and P. W. Shor, arXiv:1408.1657], 
whose ground-state can be expressed 
in terms of a uniform superposition of all colored Motzkin paths. 
Our analytical results show that for spin $s \ge 2$ 
there is a violation of the cluster decomposition property. 
This has to be contrasted with $s=1$, where the cluster property holds. 
Correspondingly, for $s=1$ one gets a light-cone profile in the propagation 
of excitations after a local quench, while the cone is absent for $s=2$,  
as shown by 
time dependent density-matrix-renormalization-group. 
Moreover, we introduce 
an original solvable model of half-integer spins 
which we refer to as Fredkin spin chain, 
whose ground-state can be expressed in terms of superposition 
of all Dyck paths. For this model we exactly calculate 
the magnetization and correlation functions,   
finding that for $s =1/2$, a cone-like 
propagation occurs while for higher spins, $s \ge 3/2$, the colors 
prevent any cone formation and clustering is violated, 
\change {together with square root deviation from the area law for the 
entanglement entropy.} 
\end{abstract}

\maketitle

\section{Introduction}
Locality plays a fundamental role in physical theories, with far reaching 
consequences, one of them being the cluster decomposition property (CDP) 
\cite{weinberg95,hastings04,sims06}. 
The CDP implies that when computing at large distances 
expectation of products of operators the product factorizes in the product 
of expectation values, therefore sufficiently distant regions behave 
independently. Of course, CDP requires that the ground-state 
is a pure state while in presence of a mixed or degenerate ground state 
the CDP may not be preserved.

Another fundamental consequence related to locality is given by the peculiar 
propagation of excitations. In particular once the system is subject to a 
local or global quench the time evolution of the correlations shows 
a well defined light cone-like propagation \cite{calabrese06}. 
For general lattice models with short-range interactions 
locality and the presence of CDP imply a bound, called Lieb-Robinson bound \cite{LRB}, 
for the commutator of two operators defined in different points of the space. 
This result is, of course, equivalent to the existence of a finite 
speed for the propagation of excitations \cite{eisert08,bloch,sotiriadis14}. This gives rise to  
a light-cone defining causally connected regions up to exponentially small 
deviations. 
\change{Actually light-cone propagation of connected correlations is 
expected when starting from an initial state with exponential clustering \cite{sims06,bravyi06,eisert06,kastner}.} 
From the other side, the presence of long-range interaction causes 
the violation of the Lieb-Robinson bound and the presence 
of power-law tails outside the light-cone 
\cite{koffel12,eisert13,hauke,metivier14,rajabpour15,carleo15}. In this 
direction 
the study of possible violations of the Lieb-Robinson bound can be experimentally performed by means of interacting trapped ions \cite{monroe}. 

The appearence of non-exponentially small corrections outside the cone 
signals non-local effects, which are induced by the long-range 
interactions or couplings. Moreover quantum correlations are also 
signaled by entanglement. This lies at the heart of 
the area law violation of the von Neumann entropy 
for long-range interacting systems. Indeed a variety of cases 
\cite{koffel12,vodola14,ramirez14,gori15,ares15}, including a study 
of the nonlinear growth after quenches \cite{growth}, have been theoretically 
investigated. 

Due to the latter arguments the study of non-local properties 
and their consequences on the light-cone propagation in addition to the 
violation of the area law are certainly at the present date a very 
challenging field of research. 
Moreover, since quantum spin chains can be used for universal quantum computation and the efficiency may be related to the amount of quantum entanglement 
\cite{vlad}, 
spin systems with more than-logarithmic entanglement entropy, as the ones we consider in this paper, can be used for quantum \changeb{computating} even more efficiently.

In this paper we intend to investigate if one can have violation of 
CDP and absence of a light-cone in the dynamics for a local quantum theory, 
which is translationally invariant in the bulk and it exhibits
a non-logarithmic violation of the area law. 
Here we first consider the exactly 
solvable chain of integer spins introduced in Ref.~\cite{bravyi12} 
for $s=1$ and recently generalized 
to larger-than-one integer spins \cite{movassagh14}. 
The peculiarity of this spin chain is that the ground-state can be expressed 
in terms of a superposition of all Motzkin paths which 
are ``colored'' for $s>1$ \cite{book}. A (non-colored) 
Motzkin path is any path 
from the point $(0,0)$ to $(0,L)$ with steps $(1,0),(1,1),(1,-1)$, {\change {where $L$ is an integer number. Any point $(x,y)$ of the path is such that $x$ and $y$ are not negative.}}
The path is said to be 
colored when the steps can be drawn with more (than one) colors. 

The following facts motivated us to investigate CDP 
and dynamics of correlations in this model: {\it i)} the model is local; 
{\it ii)} the ground-state is unique and it is a pure state made by 
a uniform superposition of all the Motzkin paths; 
{\it iii)} it exhibits a logarithmic deviation from the area law for $s=1$ and 
a square root deviation for $s \ge 2$; {\it iv)} it can be written in terms 
of spins-$s$ $S_\alpha(j)$ where $j=1,\cdots,L$ ($L$ then being interpreted as 
the number of spins) and $\alpha=x,y,z$. 
For $s=1$ the one-point {\change {ground state}} correlation functions 
were computed in Ref.~\cite{bravyi12}, while 
the two-point correlation functions were reported in Ref.~\cite{movassagh16}.

In this paper, instead, we present analytical results 
for one-point and two-point {\change {ground state}} correlation functions for any integer spin $s$, also for $s\ge 2$. In particular we will focus our attention to the connected correlation function 
\begin{equation}
\langle\langle S_z(j) S_z(k)\rangle\rangle \equiv 
\langle S_z(j) S_z(k)\rangle -\langle S_z(j) \rangle \langle S_z(k) \rangle 
\label{corr_conn}
\end{equation}
in order to see if CDP is preserved.  
Indeed, the unconnected correlation may tend, for large distances $|j-k|$, 
to a number different from zero, as it happens in presence of off-diagonal 
long-range order \cite{penrose56} (\change{or in other spin models}, 
e.g., in dimerized spin chains \cite{campos}, \change {and in AKLT model} \cite{aklt}), but the connected one goes to zero in the presence of CDP. 
  
From our analytical results we conclude that 
there is a violation of CDP: indeed, among others, 
we present a closed-form expression for  
$\langle\langle S_z(j) S_z(L-j+1)\rangle\rangle$ valid for $L \to \infty$, 
showing that it tends to a non-zero value for $s>1$, but to zero for $s=1$. 
We then correspondingly study by means of 
time-dependent density-matrix renormalization group (t-DMRG) 
\cite{feiguin2005} the dynamical 
propagation of excitations \cite{note_numerics}. 
We show that the evolution of the magnetization, once the ground state 
is perturbed by a local quench,  
exhibits a well defined light-cone profile for spin $s=1$. 
For $s=2$ instead the propagation is practically instantaneous and the cone 
formation is absent. In order to check the validity of our results we also 
calculate the magnetization spreading following the inverse path, namely by 
finding the ground state in presence of a local magnetic field and letting 
the system evolves once the latter is removed. 
Also in this configuration the colors, characteristic of the $s=2$ case, allow 
a practically instantaneous signal propagation.

In order to establish the generality of the previous results, we then 
proceed by introducing and solving a model for half-integer spins. 
This allows us to check 
if both the violation of the CDP and the absence of the light-cone 
for $s>1$ are related to the topological nature of the Motzkin paths 
and to the integerness of the spins. 
This new model, which we may refer to as the  
{\it Fredkin} model since its 
Hamiltonian can be expressed in terms of Fredkin gates \cite{nielsen}, 
has as ground state a uniform superposition of all Dyck paths \cite{book},   
as opposite to the integer case where the ground state 
is based on Motzkin paths. 
A (non-colored) Dyck path is any path 
from the point $(0,0)$ to $(0,L)$ ($L$ here should be an even integer number) 
with steps $(1,1),(1,-1)$. {\change {As for the Motzkin path, any point $(x,y)$ of the Dyck path is such that $x$ and $y$ are not negative.}}
The path is colored when the steps can be drawn with more than one 
color with the same rule as in Ref.~\cite{movassagh14}. 
Deferring details to the Appendices, one can write 
the Hamiltonian 
in terms of half-integer spins $s=1/2,3/2,\cdots$ (Appendix A) and compute 
for general $s$ the one-point and two-point correlation functions (Appendix B) 
and the dynamics after a quantum quench. Our results 
show that the von Neumann entropy exhibits just a logarithmic violation 
of the area law for $s=1/2$, while a square root violation for 
$s \ge 3/2$. Furthermore for $s=1/2$ CDP is preserved together with 
the presence of light-cone in the dynamics, while 
for $s \ge 3/2$ CDP is violated and the light-cone is absent.

\section{Spin models, magnetization and correlation functions}
In what follows we start presenting our results for the 
integer Motzkin model (both non-colored and colored) and afterwards we will 
consider the half-integer Fredkin model.

\subsection{Integer spin model (Motzkin model)}
The integer-spin Motzkin Hamiltonian 
\cite{bravyi12,movassagh14,movassagh16} can be written as 
a local Hamiltonian made by a bulk contribution, 
\bea
\nonumber H_0&=&\frac{1}{2}\sum_{c=1}^q\sum_{j=1}^{L-1}
\Big\{{\cal P}
\left(\Ket{0_j \Uparrow^{c}_{j+1}} -\Ket{\Uparrow^{c}_{j} 0_{j+1}}\right)\\
\nonumber 
&&+{\cal P}\left(\Ket{0_j \Downarrow^{c}_{j+1}} -\Ket{\Downarrow^{c}_{j} 0_{j+1}}\right)\\
&&+{\cal P}\left(\Ket{0_j 0_{j+1}} -\Ket{\Uparrow^{c}_{j} \Downarrow^c_{j+1}}\right)\Big\}
\eea
plus a crossing term 
$H_X=\sum_{c\neq \bar c}^q\sum_{j=1}^{L-1}{\cal P}\left(\Ket{\Uparrow^{c}_j 
\Downarrow^{\bar c}_{j+1}}\right)$
and a boundary term 
$H_\partial=\sum_{c=1}^q \left[{\cal P}\left(\Ket{\Downarrow^{c}_1}\right)+{\cal P}\left(\Ket{\Uparrow^{c}_L}\right)\right]$, 
where ${\cal P}(|.\rangle)$ denotes the operator $|.\rangle\langle.|$, 
and $|\Uparrow \rangle$ ($|\Downarrow \rangle$) the integer spin up (down), 
$c, \bar{c}$ 
the colors from $1$ to $q\in \mathbb{Z^+}$ and the spin $s$ 
being equal to the number of colors $q$.

The one-point and two-point correlation functions 
for $s=1$ have been computed in Ref.~\cite{bravyi12,movassagh16}. 
Here, instead, we present 
exact expressions for $\langle S_z(i)\rangle$ and 
$\langle S_z(j) S_z(k) \rangle$, 
for the general colored case, namely for any $s\ge 1$ 
(see Appendix B for further details)  
determining, in this way, the 
connected correlation functions. 
In the rest of the paper we will adopt the following notation: we 
denote expectation values by 
$\langle.\rangle_M$
for the integer spin model with the Motzkin ground state, and by 
$\langle.\rangle_D$
for the half-integer spin model with the Dyck ground state.

\begin{figure}[!ht]
\includegraphics[width=6.5cm]{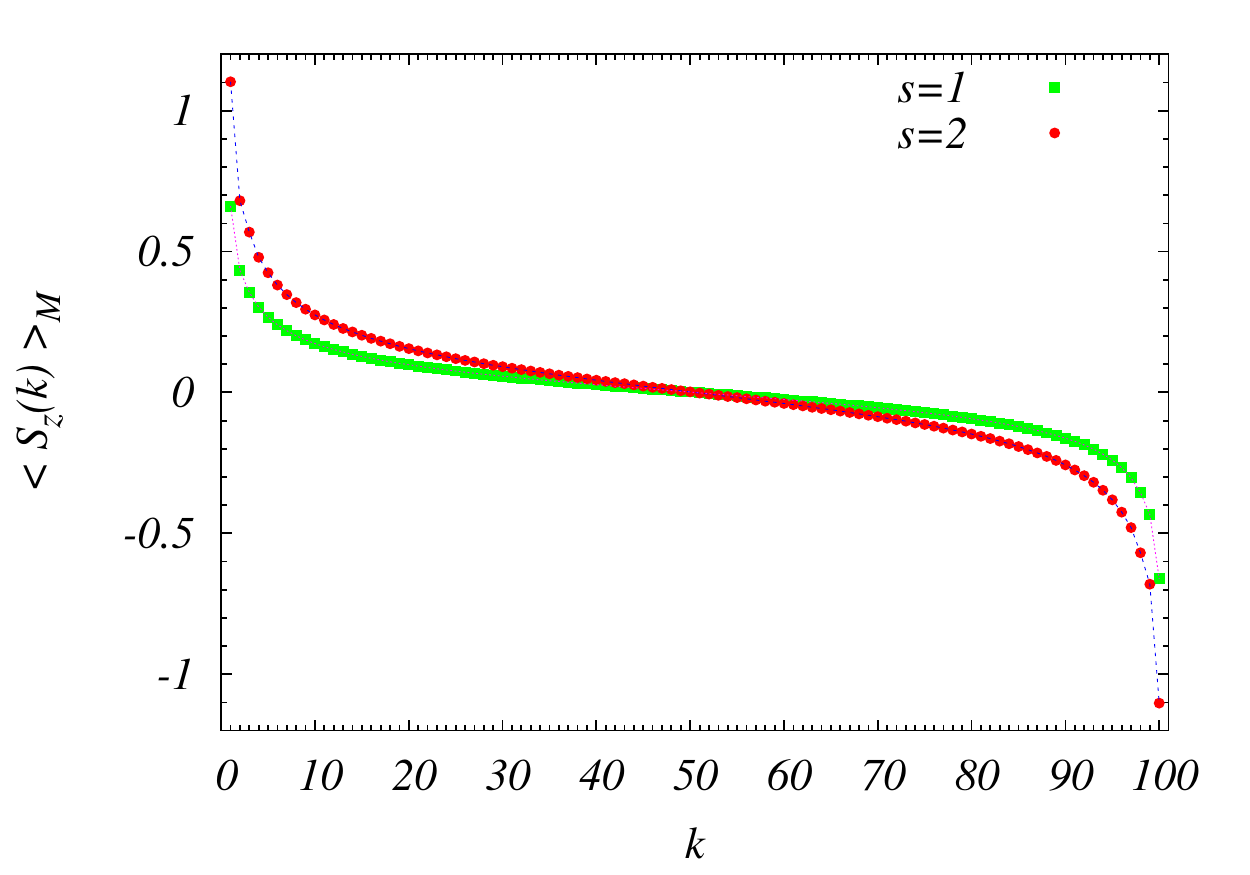}
\includegraphics[width=6.5cm]{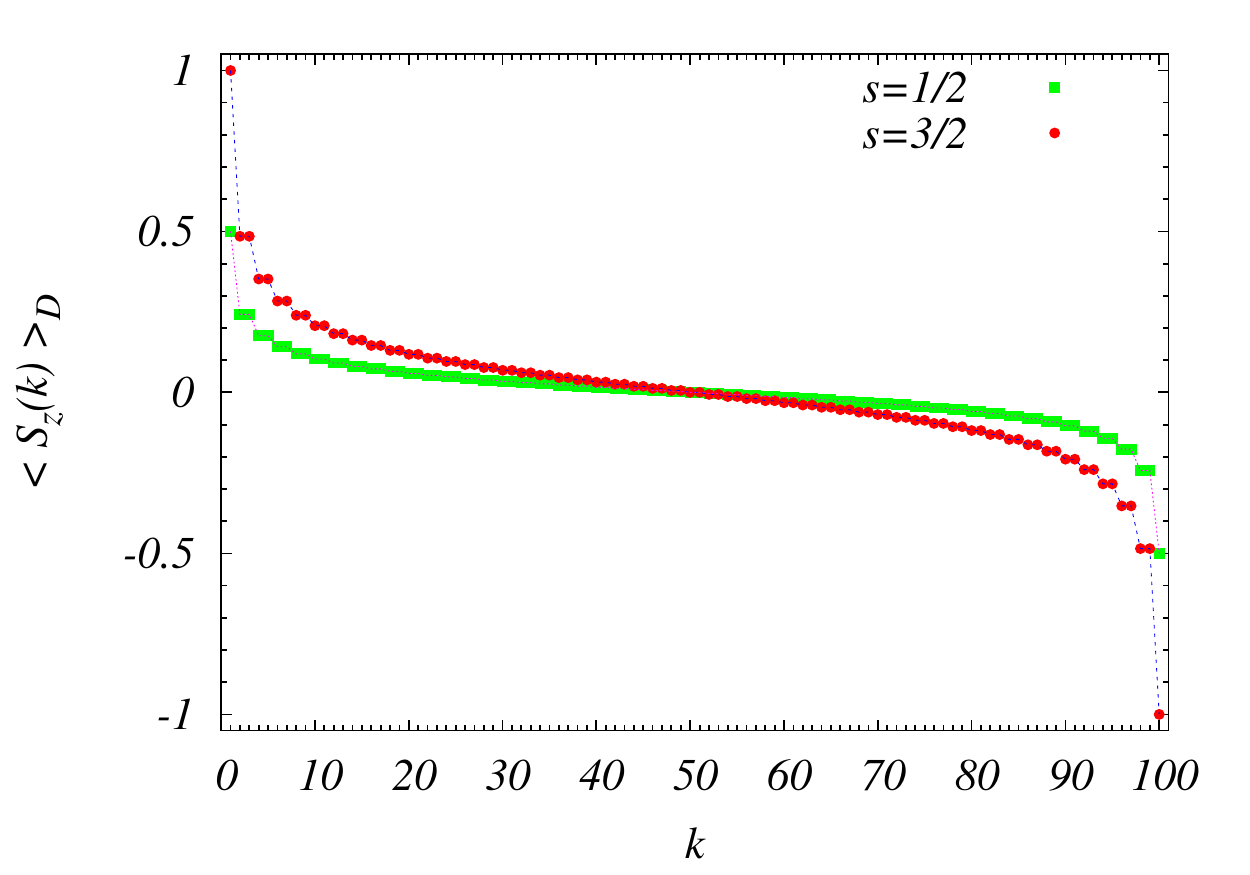}
\caption{Magnetization for integer (left) and half-integer (right) spin cases 
for a chain of lenght $L=100$.}
\label{figure1}
\end{figure}

Defining ${\cal M}^{(n)}_{hh'}$ as the number of colored Motzkin-like paths 
between two points at heights $h$ and $h'$ and linearly distant $n$ steps, 
from a combinatoric calculation 
the magnetization as a function of the position is given by
\begin{equation}
\langle S_z(j)\rangle_M
=\frac{(1+q)}{2{\cal M}^{(L)}}
\sum_{h} {\cal M}_{0 h}^{(j-1)}\left(q\,{\cal M}_{h+1, 0}^{(L-j)}-
{\cal M}_{h-1, 0}^{(L-j)}\right)
\label{analitico1}
\end{equation}
where ${\cal M}^{(L)}\equiv {\cal M}^{(L)}_{00}$ is the colored Motzkin number 
(explicit expressions are given in Appendix B). 
Equation (\ref{analitico1}), which is valid for any positive integer $s$,  
has been plotted in Fig.~\ref{figure1} (left side) and 
has been tested against Exact Diagonalization results for small sizes and 
DMRG for larger ones. As shown in Fig.~\ref{figure1}, $\langle S_z(j)\rangle$ is an odd function of the position. 
The two-point correlation function 
has been also analytically calculated and reads as it follows
\bea
\label{SzSz_int}
&&\nonumber\langle S_z(j)S_z(k)\rangle_M
=\frac{(1+q)^2}{4{\cal M}^{(L)}}
\sum_{h,h'}\,{{\cal M}}^{(j-1)}_{0h}\\
\nonumber &&\phantom{*}
\left(q\,{\cal M}_{h+1,h'}^{(k-j-1)}-{\cal M}_{h-1,h'}^{(k-j-1)}\right)\left(q\,{{\cal M}}^{(L-k)}_{h'+1,0}-{{\cal M}}^{(L-k)}_{h'-1,0}\right)\\
&&\phantom{+}
-\frac{{{\cal M}}^{(k-j-1)}}{{{\cal M}}^{(L)}}\frac{(q^3-q)}{12}
\sum_{h}{{\cal M}}^{(j-1)}_{0h}{{\cal M}}^{(L-k)}_{h0}.
\eea
The last term of the previous equation, which vanishes for $s=q=1$, 
is actually responsible for the violation of the CDP (Appendix C). 

\begin{figure}[ht!]
\includegraphics[width=6.5cm]{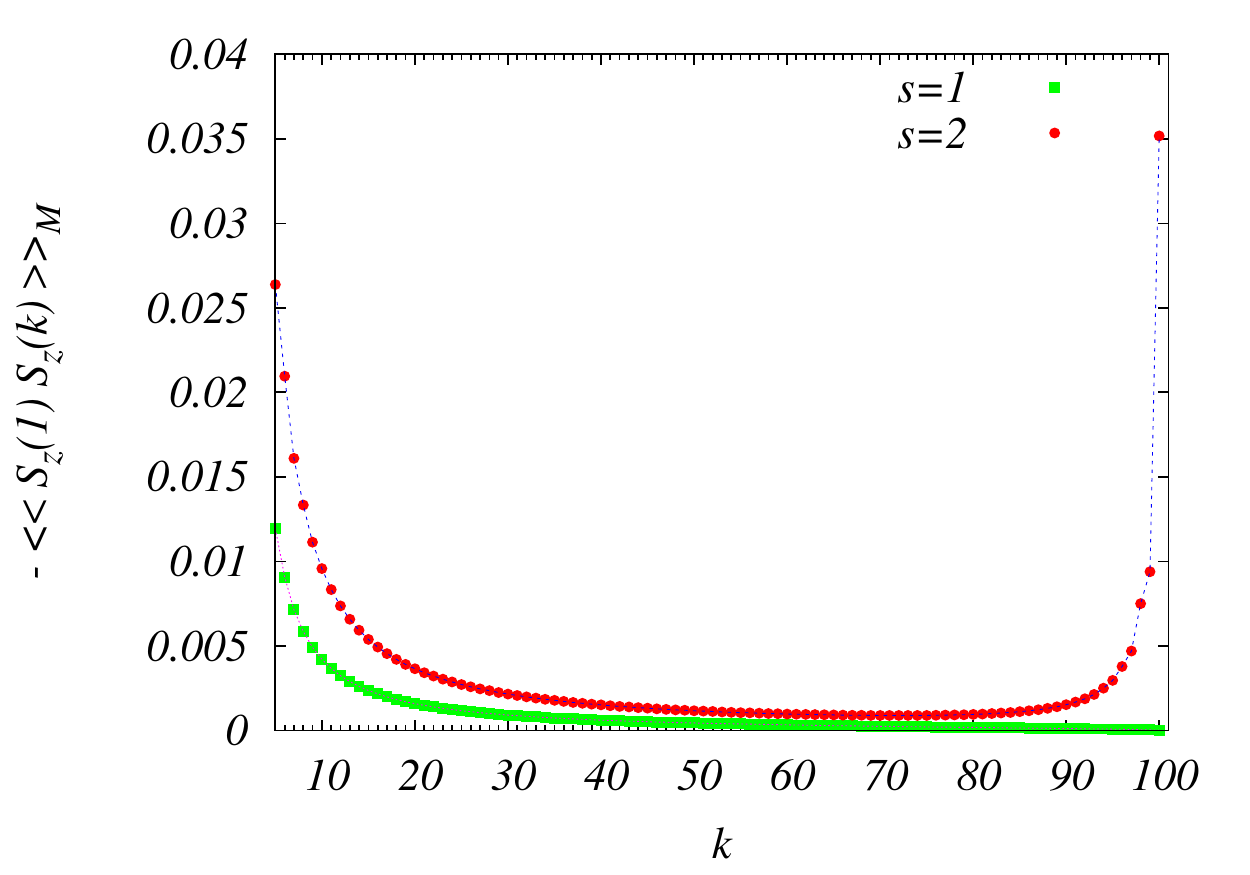}
\includegraphics[width=6.5cm]{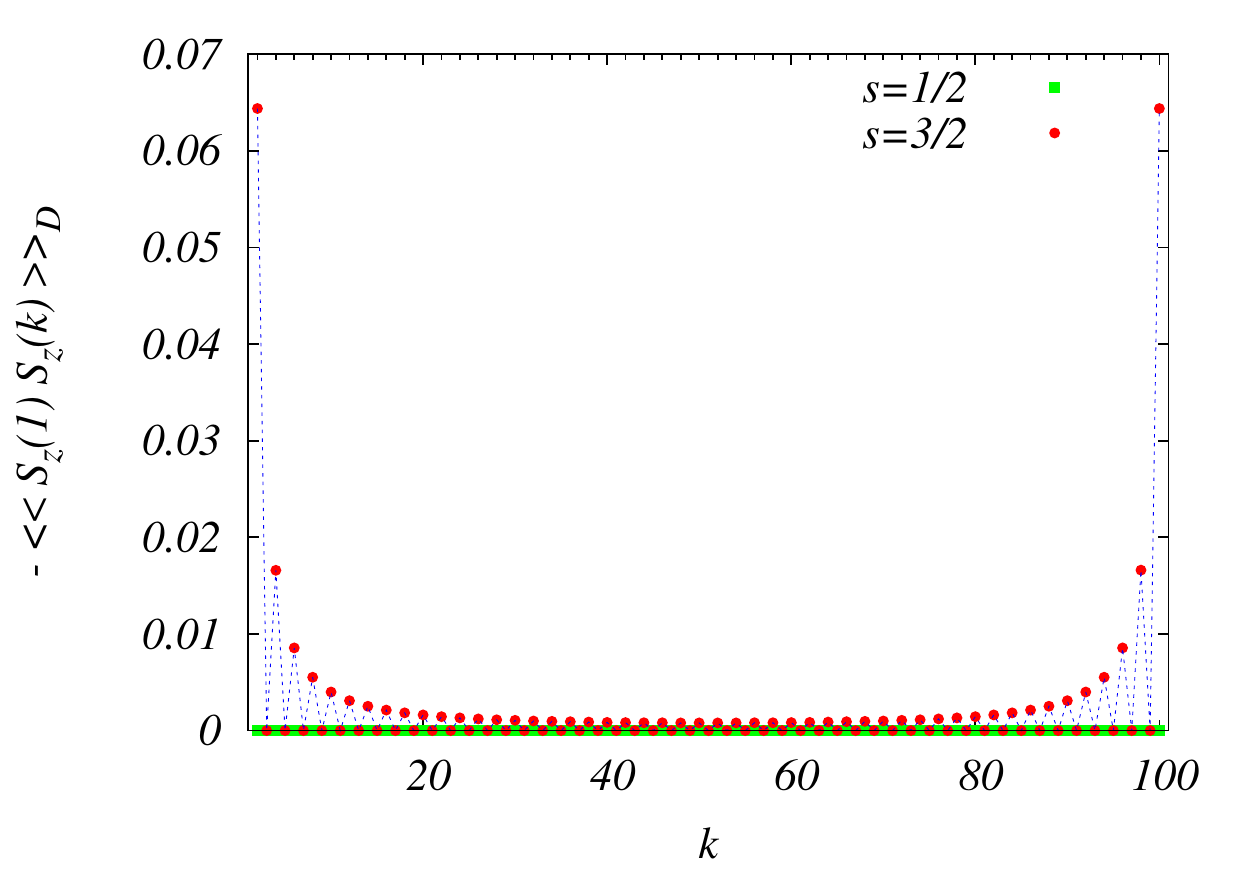}
\caption{Connected correlation functions,  
$-\langle\langle S_z(1) S_z(k) 
\rangle\rangle$, as a function of $k$ for integer (top, for $5\le k\le L$),  and half-integer 
(bottom, for $2\le k\le L$) spins for chains of length $L=100$. 
}
\label{figure2}
\end{figure}

Using Eqs.~(\ref{corr_conn}) and (\ref{SzSz_int}) we can compute  
the connected correlation function, shown for $j=1$ as a function of $k$ 
in the top part of Fig. \ref{figure2}. Analytical results in closed 
form can be obtained for the boundary connected correlation functions:
\bea
\label{S1Sk_M}
&&\hspace{-1cm}{\langle\langle} S_z(1)S_z(k)\rangle\rangle_M
\underset{k\gg 1}{\longrightarrow} \frac{(q-q^3)}{12}
\frac{{{\cal M}}^{(L-k)}{{\cal M}}^{(k-2)}}{{{\cal M}}^{(L)}}\\
&&\hspace{-1cm}{\langle\langle} S_z(j)S_z(L-j+1)\rangle\rangle_M
\underset{L\gg j}{\longrightarrow} \frac{(q-q^3)}{12}
\frac{{{\cal M}}^{(L-2j)}{{\cal M}}^{(2j-2)}}{{{\cal M}}^{(L)}}
\label{Ssym_M}
\eea
In particular for spin $s=q=2$ we get  
\be
\label{S1SL_M}
\lim_{L\rightarrow \infty} {\langle\langle} S_z(1)S_z(L)\rangle\rangle_M
=\frac{1}{2}\lim_{L\rightarrow \infty}\frac{{\cal M}^{(L-2)}}{{\cal M}^{(L)}}
\simeq -0.034
\ee
These results show that CDP is violated for colored spin chains.

\subsection{Half-integer spin model (Fredkin model)}
For the half-integer spin case, with generic spin $s=q-\frac{1}{2}$, 
we introduce the following Fredkin model 
\bea
\label{fred0}
\nonumber H_0&=&\frac{1}{2}
\sum_{c,\bar c=1}^q\Big\{\sum_{j=1}^{L-2}\Big[
{\cal P}
\left(\Ket{\downarrow^{\bar c}_j \uparrow^{c}_{j+1} \downarrow^{c}_{j+2}} -
\Ket{ \uparrow^{c}_{j} \downarrow^{c}_{j+1}\downarrow^{\bar c}_{j+2}} \right)\\
\nonumber &&+{\cal P}
\left(\Ket{\uparrow^{\bar c}_j \uparrow^{c}_{j+1} \downarrow^{c}_{j+2}} -
\Ket{\uparrow^{c}_{j} \downarrow^{c}_{j+1}\uparrow^{\bar c}_{j+2} } \right)
\Big]\\
&&+\sum_{j=1}^{L-1}{\cal P}\left(\Ket{\uparrow^{c}_j \downarrow^{c}_{j+1}} - 
\Ket{\uparrow^{\bar c}_{j} \downarrow^{\bar c}_{j+1}}\right)
\Big\}
\eea
with the inclusion of a crossing term, analogous to the previous one, 
$H_X=\sum_{c\neq \bar c}^q\sum_{j=1}^{L-1}
{\cal P}\left(\Ket{\uparrow^{c}_j \downarrow^{\bar c}_{j+1}}\right)$, 
and a boundary term
$H_\partial=\sum_{c=1}^q \left[{\cal P}\left(\Ket{\downarrow^{c}_1}\right)+
{\cal P}\left(\Ket{\uparrow^{c}_L}\right)\right]$,
where $|\uparrow^c \rangle$ is the half-integer ($c-\frac{1}{2}$)-spin up and 
$|\downarrow^c \rangle$ the half-integer ($c-\frac{1}{2}$)-spin down. A more general Hamiltonian of the Fredkin chains, having the same ground-state, but different excited states, is presented in Appendix A.

We define ${\cal D}^{(n)}_{hh'}$ the number of {\change {colored paths between two points at heights $h$ and $h'$ and linearly distant $n$, which never cross the ground}}, such that ${\cal D}^{(2n)}\equiv{\cal D}^{(2n)}_{00}=q^{n}C(n)$, where $C(n)$ are the Catalan numbers (see Appendix B). 
The magnetizations is, then, found to be 
\begin{equation}
\langle S_z(j)\rangle_D
=\frac{q}{2{\cal D}^{(L)}}
\sum_{h} {\cal D}_{0 h}^{(j-1)}\left(q\,{\cal D}_{h+1, 0}^{(L-j)}-{\cal D}_{h-1, 0}^{(L-j)}\right)
\label{analitico1_Fredkin}
\end{equation}
(the subscript $_D$ 
refers to Dyck ground state). 
The result Eq.~(\ref{analitico1_Fredkin}) is plotted 
in the bottom part of Fig. \ref{figure1}, for spins $s=1/2$ and $s=3/2$.
We can also calculate analytically the correlation functions getting 
\bea
\label{SzSz_hint}
&&\nonumber\langle S_z(j)S_z(k)\rangle_D
=\frac{q^2}{4{\cal D}^{(L)}}
\sum_{h,h'}\,{{\cal D}}^{(j-1)}_{0h}\\
\nonumber &&\phantom{*}
\left(q\,{\cal D}_{h+1,h'}^{(k-j-1)}-{\cal D}_{h-1,h'}^{(k-j-1)}\right)\left(q\,{{\cal D}}^{(L-k)}_{h'+1,0}-{{\cal D}}^{(L-k)}_{h'-1,0}\right)\\
&&\phantom{+}
-\frac{{{\cal D}}^{(k-j-1)}}{{{\cal D}}^{(L)}}\frac{(q^3-q)}{12}
\sum_{h}{{\cal D}}^{(j-1)}_{0h}{{\cal D}}^{(L-k)}_{h0}
\eea
Also in this case the last term which vanishes for $q=1$, 
i.e. $s=1/2$, is responsable for the violation of the cluster decomposition 
property. 
In particular, for $j=1$, and $1<k\le L$,  we get simply 
\be
{\langle\langle} S_z(1)S_z(k)\rangle\rangle_D
=\frac{(1-q^2)}{12} p_k \frac{C(\frac{L-k}{2})C(\frac{k}{2}-1)}{C(\frac{L}{2})}
\label{S1Sk_D}
\ee
which for $q=1$ (spin $s=1/2$) is exactly zero, since the spin at the first 
site has to be $\uparrow$, no matter the rest of the chain. In Eq.~(\ref{S1Sk_D}) $p_k=1$ for even $k$ and $p_k=0$ for odd $k$. 
Putting $k=L$ and sending $L\rightarrow \infty$ we have, however, a finite 
correlation
\be                                                                            
\lim_{L\rightarrow \infty} {\langle\langle} S_z(1)S_z(L)\rangle\rangle_D
=\frac{(1-q^2)}{12} \frac{1}{4}.
\label{S1SL}
\ee 
More generally, the analytic expression for the correlators, in the long 
$L$ limit, and $j\ge 1$, are  
\bea
\label{SjSL}
\hspace{-1cm}&&
\lim_{L\rightarrow \infty} {\langle\langle} S_z(j)S_z(L)\rangle\rangle_D
=\frac{(1-q^2)}{12} \frac{C(\frac{j-1}{2})}{2^{j+1}} p_{(j+1)}\\
\hspace{-1cm}&&
\lim_{L\rightarrow \infty} {\langle\langle} S_z(j)S_z(L-j+1)\rangle\rangle_D
=\frac{(1-q^2)}{12} \frac{C(j-1)}{4^j}
\label{Ssym}
\eea
We show therefore that, also in the half-integer spin chains, for $s\ge 3/2$ 
the violation of the CDP occurs. 
The connected correlation functions for $s=1/2$ and $s=3/2$ 
are plotted in the bottom part of Fig.~\ref{figure2}. 
\change{In Appendix 
we plot for comparison static DMRG results for {\it i}) 
Heisenberg $XXX$ Hamiltonian with 
boundary magnetic term $H_\partial$, for both spin $1/2$ and $3/2$; 
{\it ii}) AKLT model with bondary term $H_\partial$, for spin $s=1$.
The result is that in all these models CDP is preserved.}
%

\section{Dynamics}

\begin{figure}[ht!]
\includegraphics[width=3.5cm]{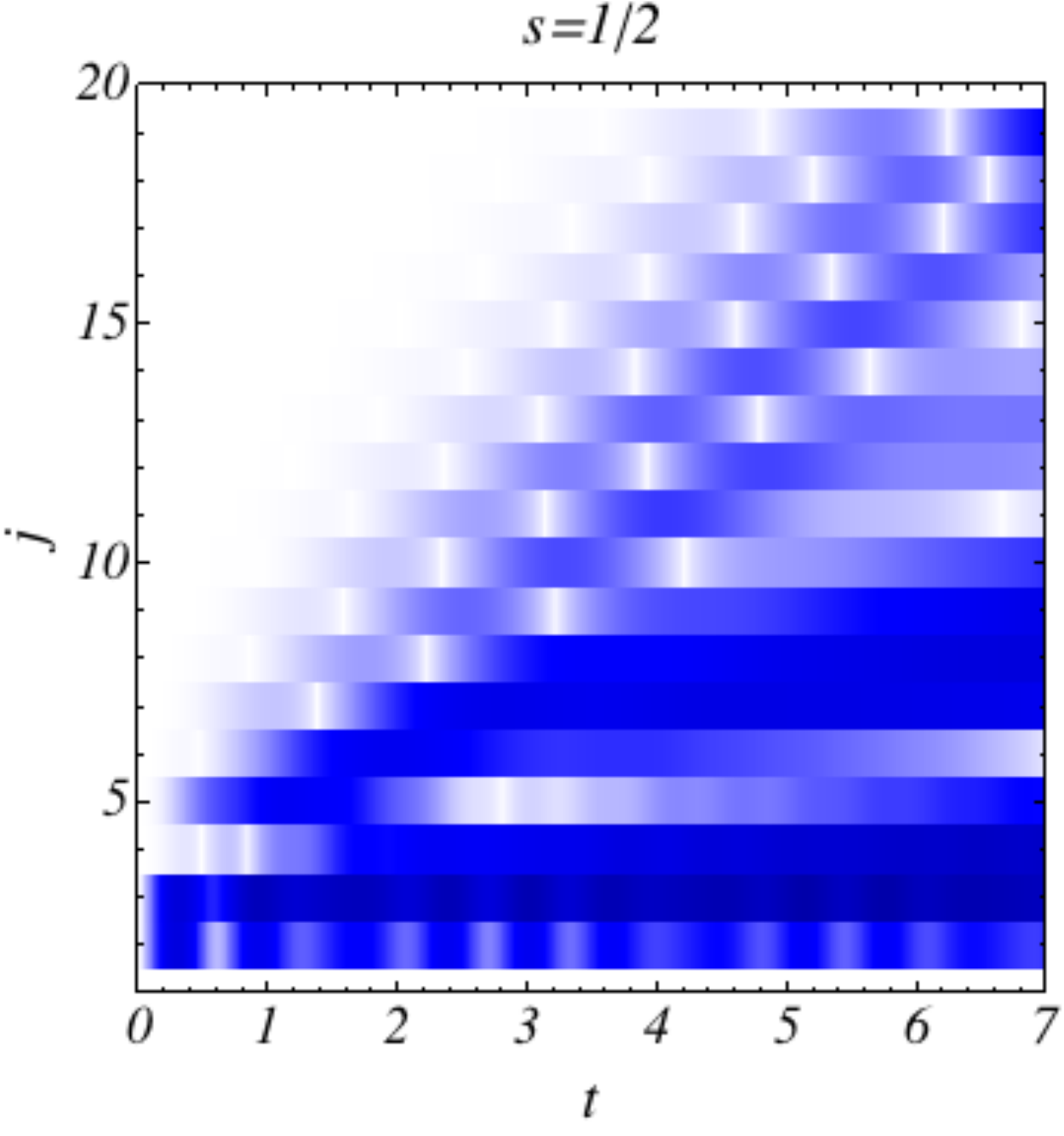}
\includegraphics[width=3.5cm]{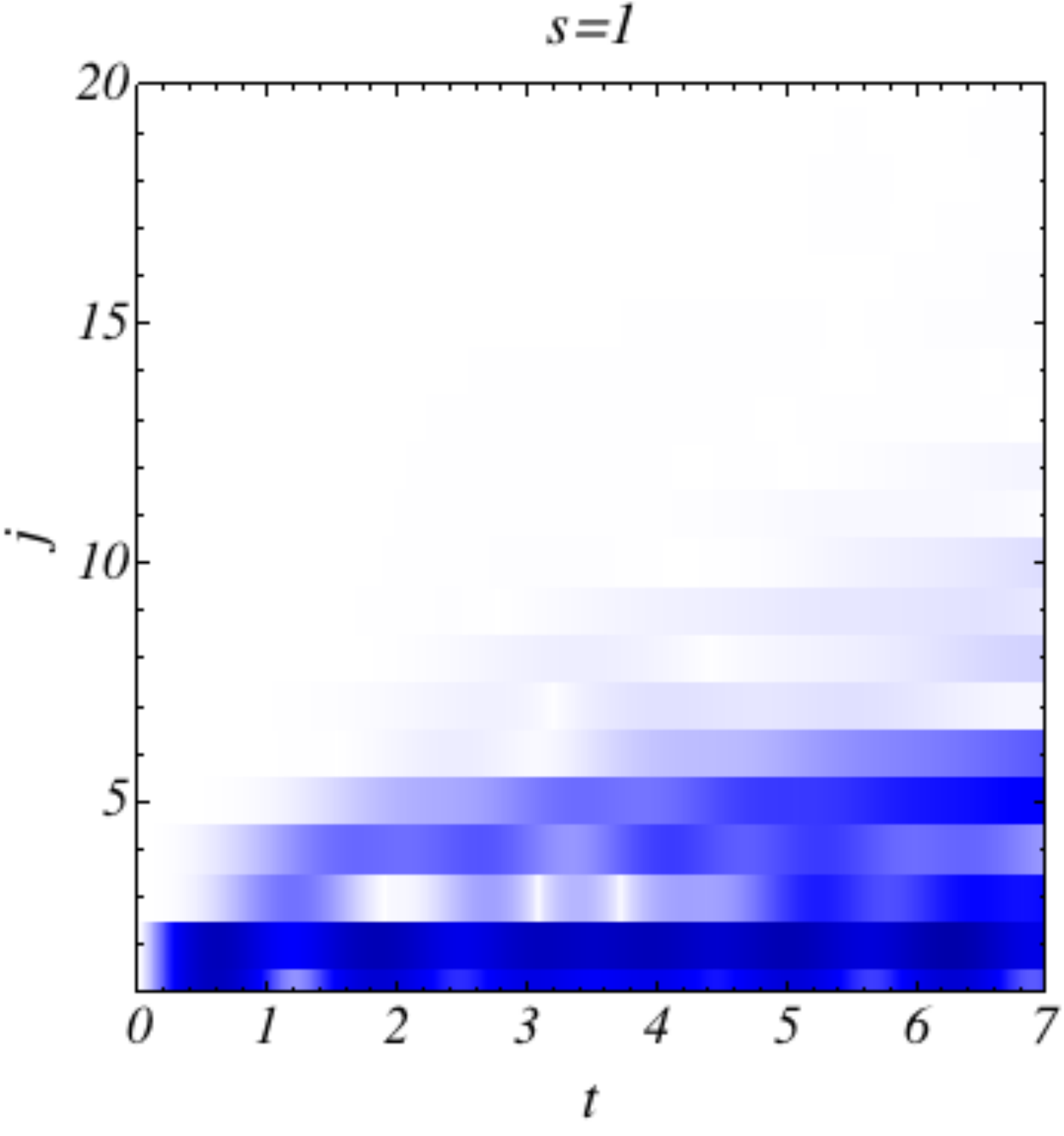}
\includegraphics[width=3.5cm]{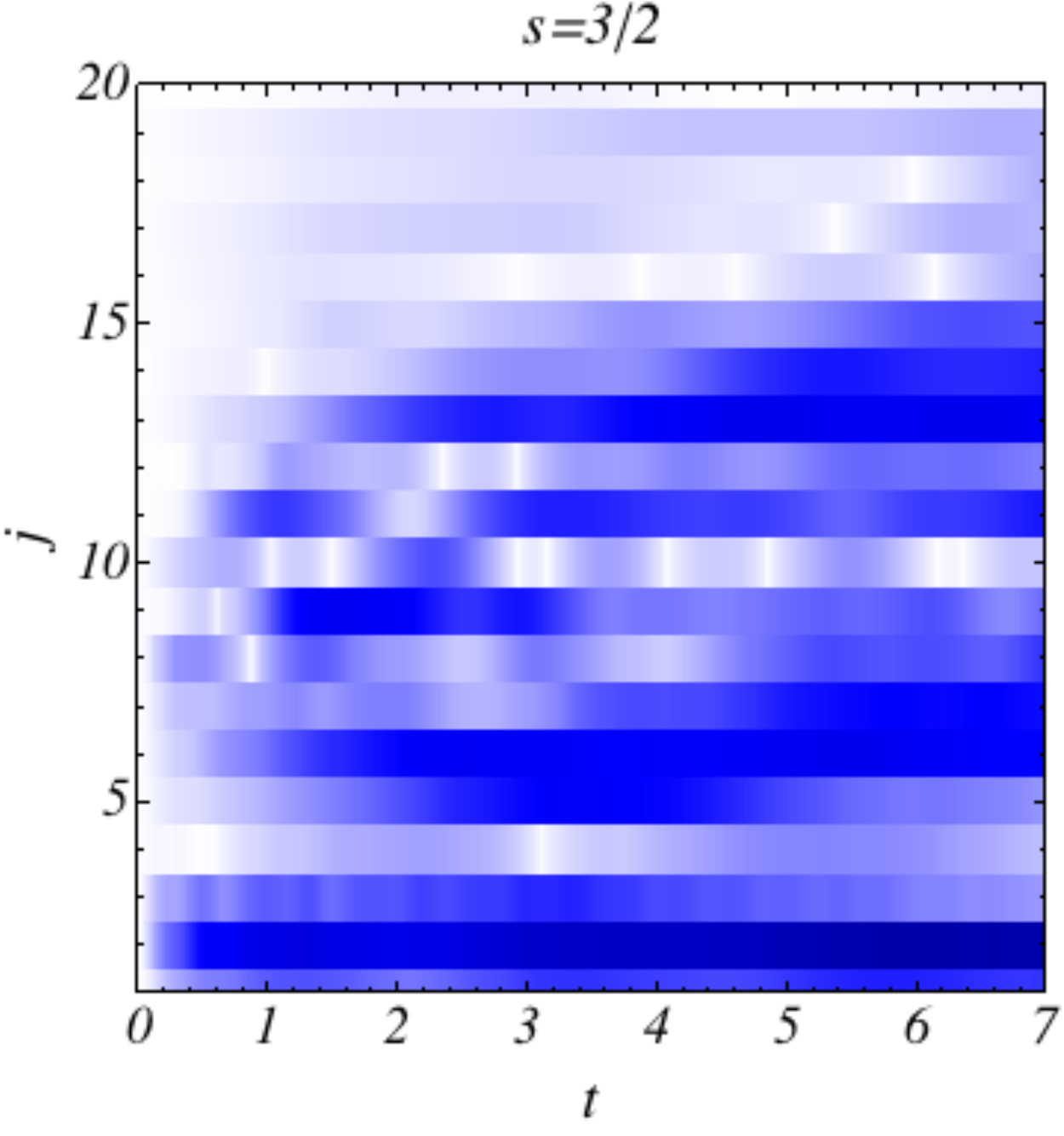}
\includegraphics[width=3.5cm]{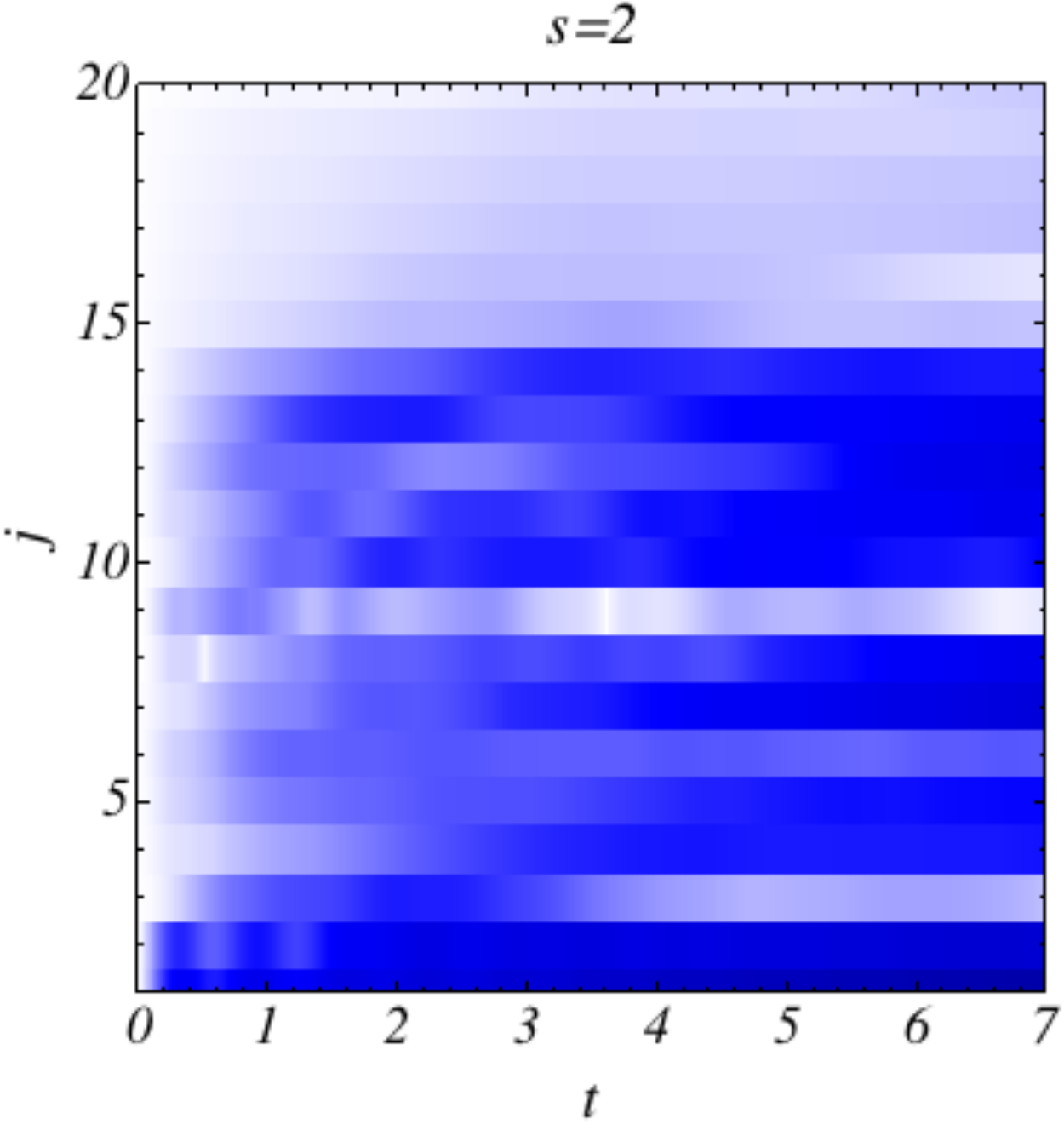}
\caption{Time evolution of $\langle S_z(j,t)\rangle-\langle S_z(j,0)\rangle$
after switching on a local field, $5\,S_z(j_0)$: a) on $j_0=2$ of a spin
$s=1/2$ chain, b) on $j_0=1$ of a spin $s=1$ chain,
c) on $j_0=2$ of a spin $s=3/2$ chain, d) on $j_0=1$ of a spin $s=2$ chain.
The light-cones are absent for $s=3/2$ and $s=2$.}
\label{figure3}
\end{figure}

\changeb{Motivated by the fact that CDP violation could give rise to relevant dynamical properties \cite{sims06,bravyi06,eisert06,kastner}, we study, by means of t-DMRG, the time evolution of the previous models once a local quench is performed.} 
As we notice 
in Fig. \ref{figure3}, once the colors are present, both for ingeger and half-integer cases, the system does not exibit light-cone propagation. 
On the contrary, for the single-colored cases ($s=1/2$ and $s=1$) a clear signature of light-cone is visible in the evolution of local 
magnetization, after switching on a local field. 

\begin{figure}[ht!]
\includegraphics[width=3.5cm]{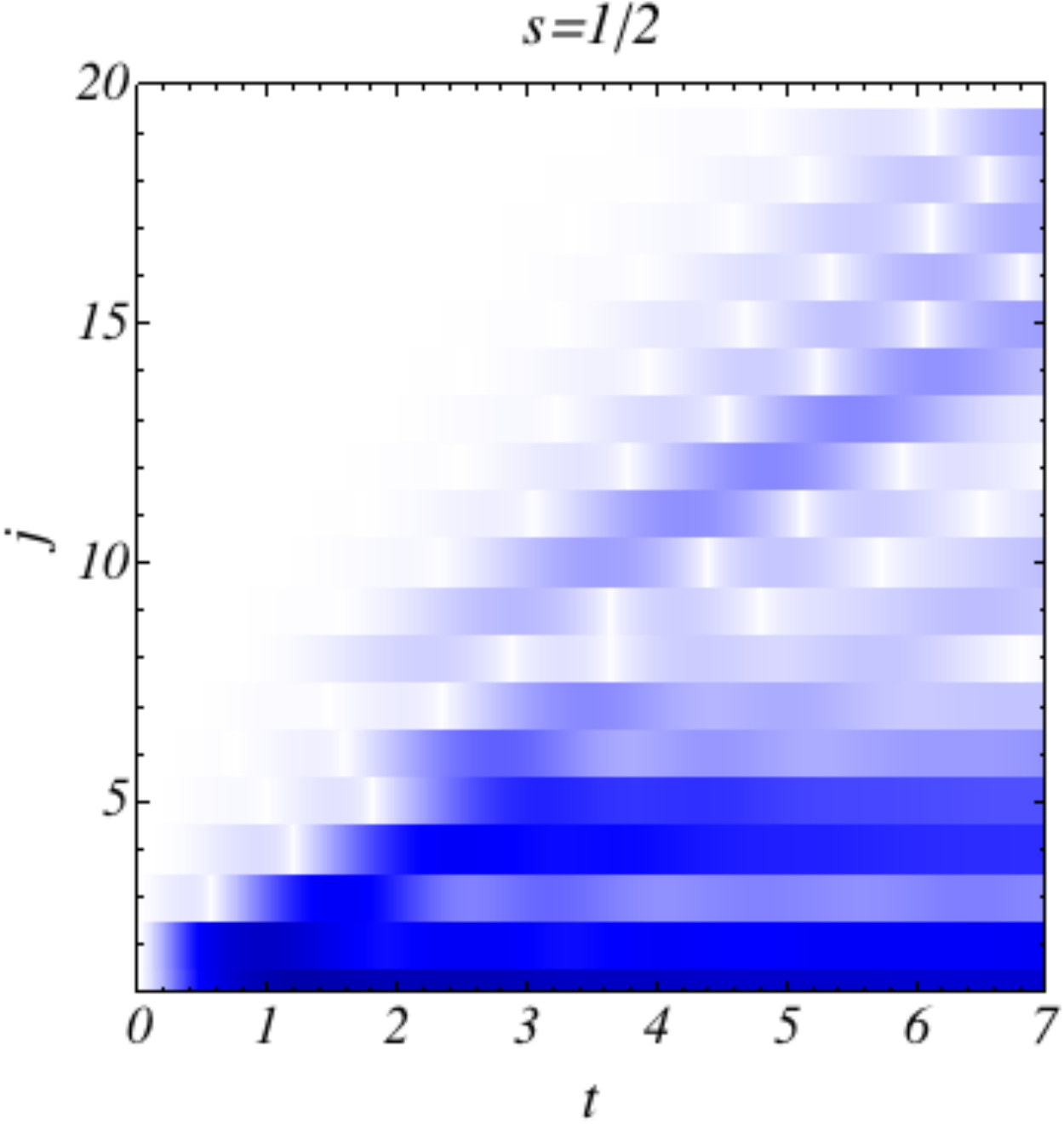}
\includegraphics[width=3.5cm]{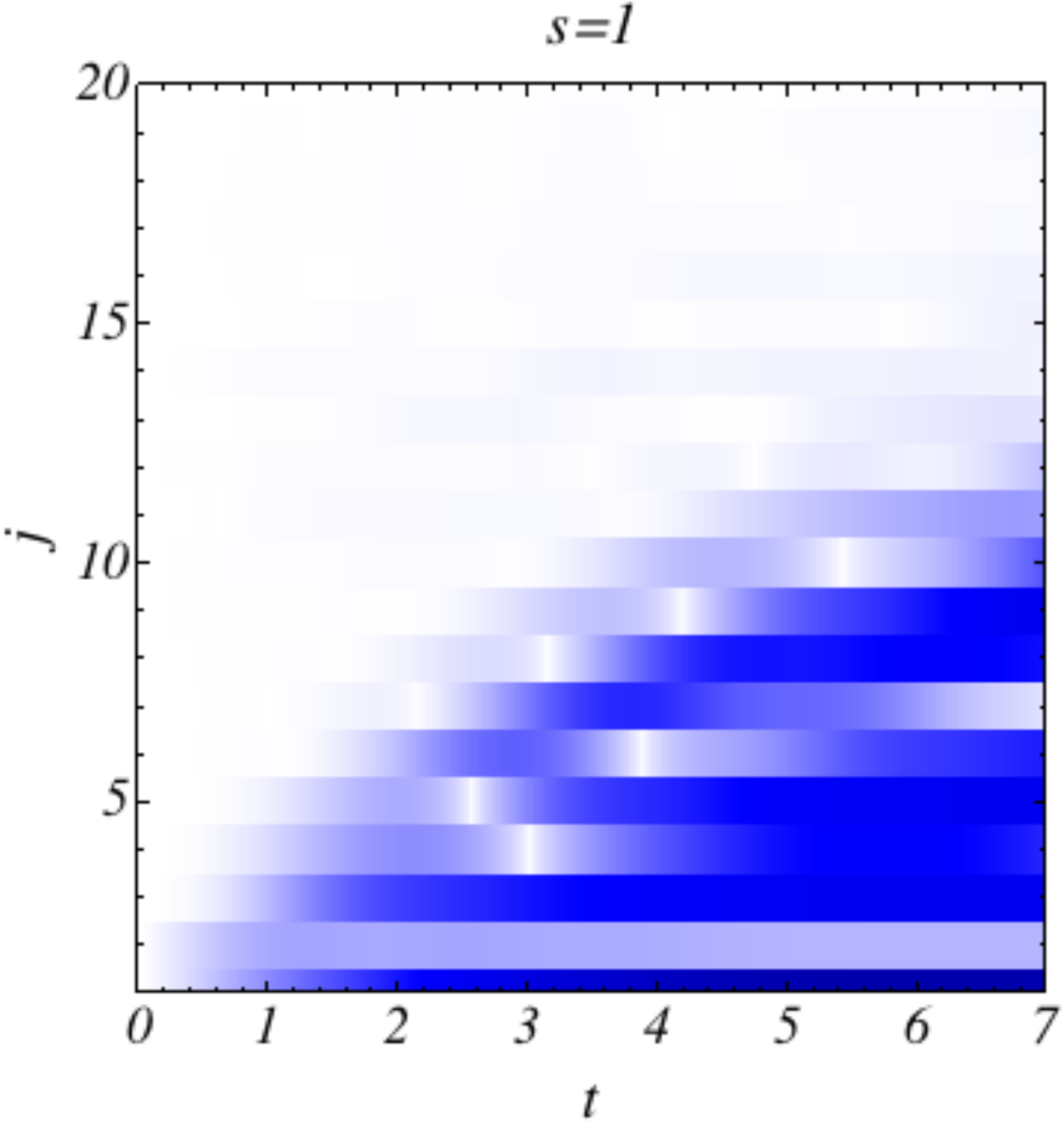}
\includegraphics[width=3.5cm]{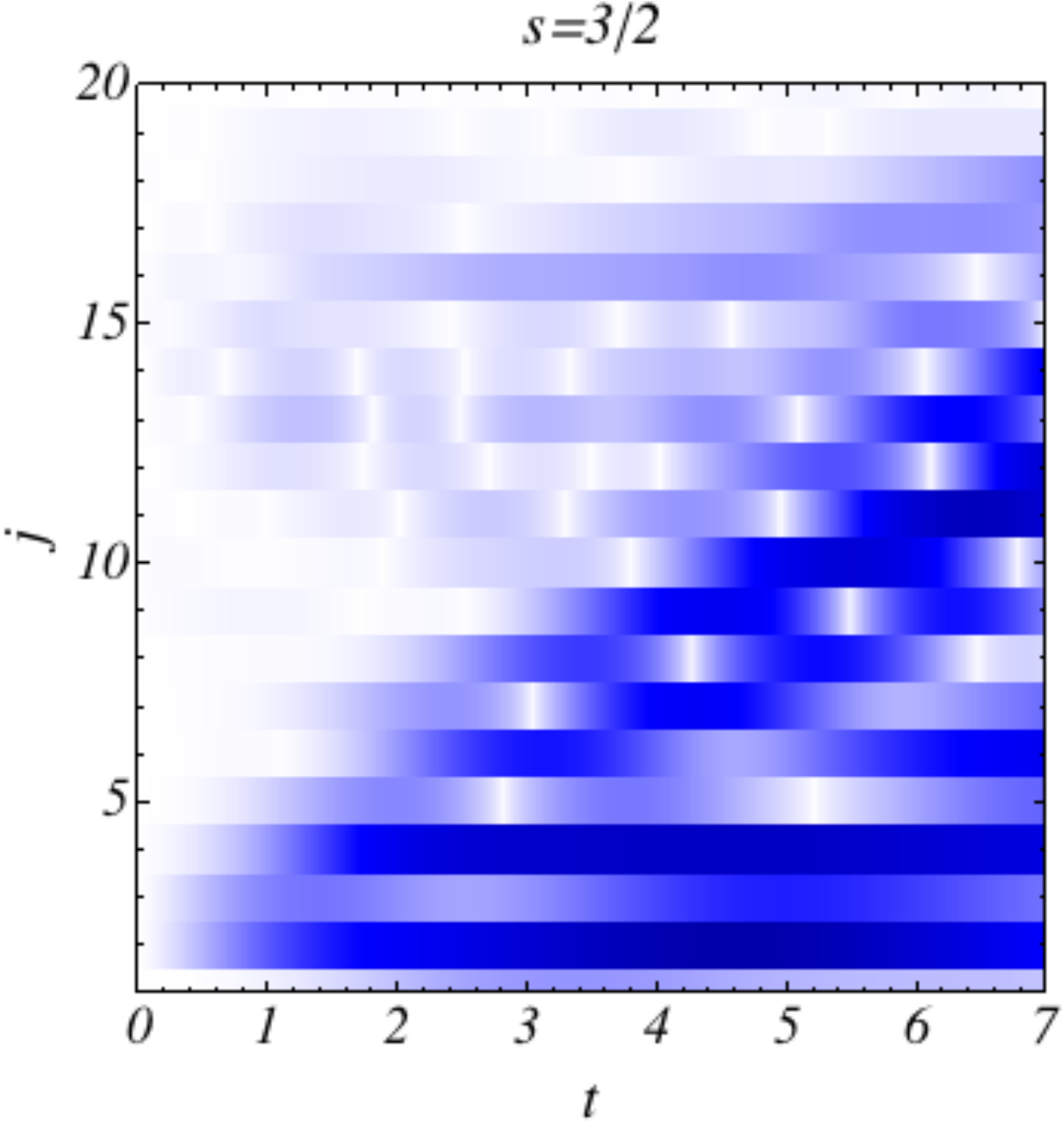}
\includegraphics[width=3.5cm]{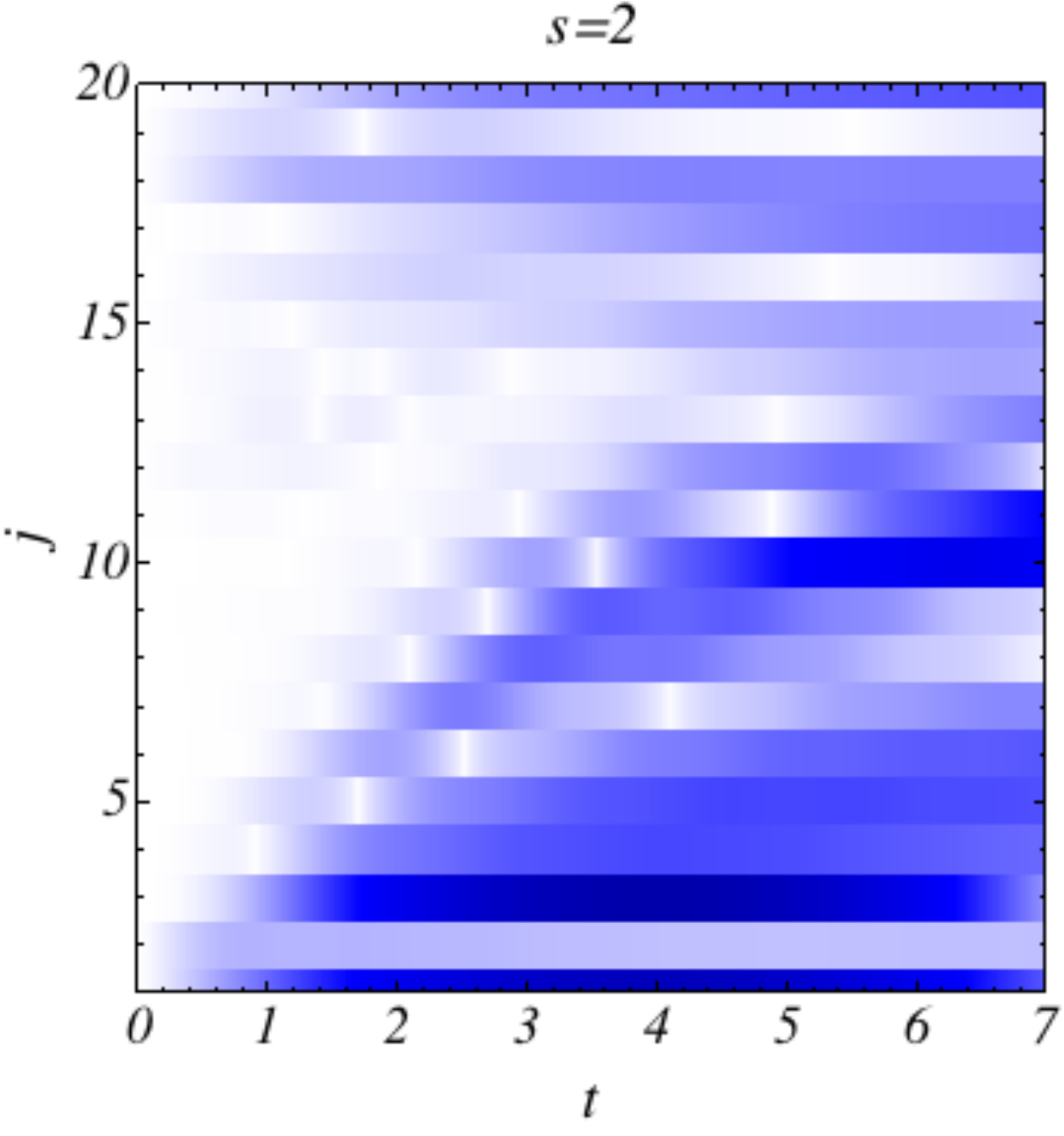}
\caption{Time evolution of $\langle S_z(j,t)\rangle-\langle S_z(j,0)\rangle$
after switching off a local field, $5\,S_z(j_0)$: a) on $j_0=2$ of a spin
$s=1/2$ chain, b) on $j_0=1$ of a spin $s=1$ chain,
c) on $j_0=2$ of a spin $s=3/2$ chain, d) on $j_0=1$ of a spin $s=2$ chain.
The cones are absent for $s=3/2, 2$.}
\label{figure2app}
\end{figure}

In Fig.~\ref{figure2app} we report the results for the time evolution of the magnetization after switching off a local field placed close to the edge of the spin chain.
Also in this case (as that shown in the main text where a local field is switched on) the light-cone is present for $s=1/2$ and $s=1$ while it is absent for $s=3/2$ and $s=2$. 
We verified that also the connected $z$-$z$
correlation functions exhibith a similar behavior.

\section{Entanglement entropy}

Finally, we report for completeness general 
results for the von Neumann entropy. 
For integer spins the area law is violated \cite{movassagh14}. 
We find that for both integer and half-integer chains the entanglement entropy 
do not follow the area law. 
We find that the entropies for the Motzkin and the Fredkin models, 
after a bipartition of the chain in two parts, $[1,j]$ and $[j+1,L]$, 
are gven by 
\bea
&&{\cal S}_M
=\log_2(q)\,\langle h_j\rangle_M
+O(\log_2(j(L-j)/L))\\
&&{\cal S}_D
=\log_2(q)\,\langle h_j\rangle_D
+O(\log_2(j(L-j)/L))
\eea
namely, the leading contrbution, for colored cases ($q>1$) is no longer 
a logarithmic term (Appendix D). 
Remarkably we show that, in both cases, the entanglement entropy 
has a geometrical meaning, since  
the leading term is proportional to the 
average height of the Motzkin paths (for the integer case) and 
Dyck paths
(for the half-integer case), measured right at the bipartition position $j$. 
This quantity, in both cases, 
is approximatelly given by 
\be
\langle h_j\rangle \approx \sqrt{\frac{2j\left(L-j\right)}{L}}.
\ee

\section{Scaling of the energy gap}
To conclude, we also report results for the finite-size scaling 
of the gap $\Delta E=E_1-E_{GS}$ where $E_{GS}=0$ is the \changeb{ground-state energy} and $E_1$ the energy of the first excited state. These energies are obtained by performing static DMRG simulations of systems with $L$ up to 100 for $s=1/2$ and $s=1$ and up to $L=60$ for $s=3/2,\,2$. We kept at most 2000 DMRG states and 5 finite size sweeps. We found that the gap scales as $\Delta \propto 1/L^c$ and, for those four spins, we always get  $c>2$. In particular our estimates are 
$c=2.9 \pm 0.1$ ($s=1/2$), $c=2.7 \pm 0.1$ ($s=1$), 
$c=3.8 \pm 0.4$ ($s=3/2$) and $c=3.3 \pm 0.3$ ($s=2$).
We think that the analytic expression for $c$ is different for Motzkin and 
Fredkin cases, still we conjecture that in both cases $c$ increases 
linearly with $\log q$.

\section{Conclusions} 
We exactly computed the magnetization and the connected 
$z$-$z$ correlation functions of a local quantum spin chain, for 
any integer spins, whose ground-state 
can be represented by 
a uniform superposition of all 
colored Motzkin paths. 
Our analytical results show that, except for $s=1$, for any $s \ge 2$ 
there is violation of 
CDP 
and that the connected correlation function 
$\langle\langle S_z(j) S_z(L-j+1) \rangle \rangle$ tends 
to a finite value for $L \to \infty$. 
Motivated by the violation of the clustering, 
we studied 
the dynamics of magnetization 
and 
correlation functions after a quench. We showed that 
for $s=1$ one has a light-cone for \changeb{the exitation propagations},  
while the cone is absent for $s=2$. 
We also introduced 
another solvable 
model of half-integer spins, that we called Fredkin 
chain, 
whose ground-state 
is 
expressed in terms of uniform superposition 
of colored Dyck paths.
We exactly computed the magnetization and the connected $z$-$z$ correlation 
functions, finding that also in this case CDP
holds for $s=1/2$, while is violated 
for $s \ge 3/2$. 
Analogously to the integer spin case, 
t-DMRG 
indicates that there is a light-cone for $s=1/2$, 
while the cone 
is absent for $s=3/2$. 
We finally computed the von Neumann entropy of the Fredkin model, 
showing a (non-logarithmic) square-root violation of the area law, 
and the scaling behavior of the first gap.
To conclude we observe that it would be interesting studying the quantum transfer of states \cite{Lorenzo} via the colored Motzkin and Fredkin spin chains considering as sender and receiver the two spins at the edges. It is also worthwhile for future investigations to systematically compare the dynamical properties of spin models considered in this paper with those of other spin chains where a global constraint is added to Hamiltonians with local interactions. 


\begin{acknowledgments}
{\it Acknowledgments:} We thank F. Franchini, L. Lepori and S. Paganelli for discussions. LD thanks SISSA for kind hospitality. LD and LB
 acknowledge financial support from MIUR through FIRB Project 
No. RBFR12NLNA\_002. LB acknowledges BEC Center in Trento and CNR-IOM Karma cluster in Trieste for CPU time.  
VEK and OS are grateful to Simons Center for Geometry and Physics for support and hospitality.
\end{acknowledgments}

\appendix
\section{Models and ground states}

The Hamiltonians we consider can be written in the form
\be
H=H_0+H_X+H_{\partial}.
\ee
For the integer spin model $(s=q)$ we define: 
$\Ket{\Uparrow^c}=\Ket{c}$, $\Ket{\Downarrow^c}=\Ket{-c}$ with $c=1,2,...,q$, 
where $q\in \mathbb{Z^+}$
\bea
H_0&=&\frac{1}{2}\sum_{c=1}^q\sum_{j=1}^{L-1}
\Big\{\\
&&\nonumber 
\left(\Ket{0_j \Uparrow^{c}_{j+1}} -\Ket{\Uparrow^{c}_{j} 0_{j+1}}\right)
\left(\Bra{0_j \Uparrow^{c}_{j+1}} -\Bra{\Uparrow^{c}_{j} 0_{j+1}}\right)\\
\nonumber
&+&\left(\Ket{0_j \Downarrow^{c}_{j+1}} -\Ket{\Downarrow^{c}_{j} 0_{j+1}}\right)
\left(\Bra{0_j \Downarrow^{c}_{j+1}} -\Bra{\Downarrow^{c}_{j} 0_{j+1}}\right)\\
\nonumber
&+&\left(\Ket{0_j 0_{j+1}} -\Ket{\Uparrow^{c}_{j} \Downarrow^c_{j+1}}\right)
\left(\Bra{0_j 0_{j+1}} -\Bra{\Uparrow^{c}_{j} \Downarrow^c_{j+1}}\right)\Big\}\\
H_X&=&\sum_{c\neq \bar{c}}^q \sum_{j=1}^{L-1}
\Ket{\Uparrow^{c}_{j}\Downarrow^{\bar c}_{j+1}}\Bra{\Uparrow^{c}_{j}\Downarrow^{\bar c}_{j+1}}\\
H_\partial&=&\sum_{c=1}^q\left(\Ket{\Downarrow^c_1}\Bra{\Downarrow^c_1}+\Ket{\Uparrow^c_L}\Bra{\Uparrow^c_L}\right).
\eea

For the half-integer spin model $\left(s=q-\frac{1}{2}\right)$ we define: 
$\Ket{\uparrow^c}=\Ket{c-\frac{1}{2}}$, 
$\Ket{\downarrow^c}=\Ket{\frac{1}{2}-c}$ with $c=1,2,...,q$ and 
\bea
\label{fred}
H_0&=&\frac{1}{2}
\sum_{c,\bar c=1}^q\Big\{\sum_{c'=1}^q\sum_{j=1}^{L-2}\\
\nonumber &&\Big[
\left(\Ket{\downarrow^{\bar c}_j \uparrow^{c}_{j+1} \downarrow^{c'}_{j+2}} -\Ket{ \uparrow^{c}_{j} \downarrow^{c'}_{j+1} 
\downarrow^{\bar c}_{j+2}} \right)\\
\nonumber
&&\left(\Bra{\downarrow^{\bar c}_j \uparrow^{c}_{j+1} \downarrow^{c'}_{j+2}} -\Bra{ \uparrow^{c}_{j} \downarrow^{c'}_{j+1} 
\downarrow^{\bar c}_{j+2}} \right) \\
\nonumber
&+&\left(\Ket{\uparrow^{\bar c}_j \uparrow^{c}_{j+1} \downarrow^{c'}_{j+2}} -
\Ket{\uparrow^{c}_{j} \downarrow^{c'}_{j+1}\uparrow^{\bar c}_{j+2} } \right)\\
\nonumber
&&\left(\Bra{\uparrow^{\bar c}_j \uparrow^{c}_{j+1} \downarrow^{c'}_{j+2}} -
\Bra{\uparrow^{c}_{j} \downarrow^{c'}_{j+1}\uparrow^{\bar c}_{j+2} } \right)\Big]\\
\nonumber
&+&\sum_{j=1}^{L-1}\left(\Ket{\uparrow^{c}_j \downarrow^{c}_{j+1}} - \Ket{\uparrow^{\bar c}_{j} \downarrow^{\bar c}_{j+1}}\right)\left(\Bra{\uparrow^{c}_{j} \downarrow^{c}_{j+1}} - \Bra{\uparrow^{\bar c}_j \downarrow^{\bar c}_{j+1}}\right)\Big\}\\
H_X&=&\sum_{c\neq \bar c}^q\sum_{j=1}^{L-1}\Ket{\uparrow^{c}_j \downarrow^{\bar c}_{j+1}}\Bra{\uparrow^{c}_j \downarrow^{\bar c}_{j+1}}\\
H_\partial&=&\sum_{c=1}^q \left(\Ket{\downarrow^{c}_1}\Bra{\downarrow^{c}_1}+\Ket{\uparrow^{c}_L}\Bra{\uparrow^{c}_L} \right)
\eea
A simplified half-integer spin model with the same ground state but which requires less computational effort, can be obtained from Eq.~(\ref{fred}), considering only the terms with $c'=c$, as in Eq.~(\ref{fred0}). 

The ground states of these frustration-free Hamiltonians are unique, made by simple superpositions of all Motzkin paths for the integer case and all Dyck paths for the half-integer one. 
Denoting $\Ket{\Uparrow}$ by $\Ket{/}$, $\Ket{\Downarrow}$ by 
$\Ket{\backslash}$ 
and $\Ket{0}$ by $\Ket{-}$ one can construct a Motzkin path ($\Ket{m}$), 
while by using only $\Ket{/}$ and $\Ket{\backslash}$ one can construct a Dick 
path ($\Ket{d}$). 
For colored paths, the colors are such that they match for up-down couples of 
spins at any height. 
\begin{figure}[h]
\includegraphics[height=2cm]{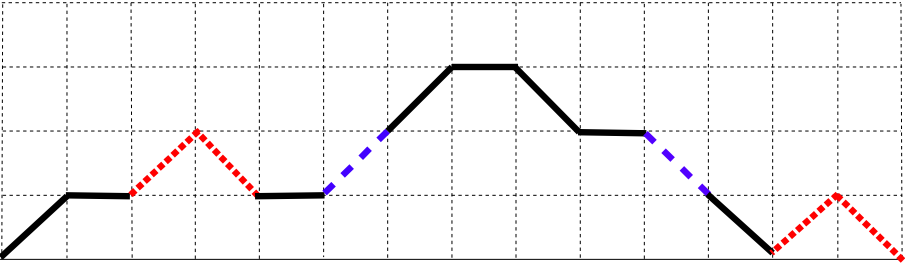}\\
\includegraphics[height=2cm]{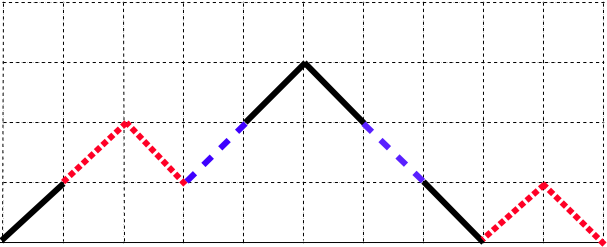}
\caption{a) Example of a colored Motzkin path, with $q=3$ colors, 
which corresponds to a particular Motzkin spin state 
$\Ket{m^{(14)}_p}=\Ket{\Uparrow^1_1 0_2\Uparrow^2_3\,\Downarrow^2_4 0_5\Uparrow_6^3\,\Uparrow_7^1 0_8 \Downarrow_9^1 0_{10} \Downarrow_{11}^3 \,\Downarrow_{12}^1 \,\Uparrow^2_{13} \,\Downarrow^2_{14}}$, b) Example of a colored Dyck 
path which corresponds to a particular Dyck spin state
$\Ket{d^{(10)}_p}=\Ket{\uparrow^1_1 \,\uparrow^2_2\,\downarrow^2_3 \uparrow_4^3\,\uparrow_5^1 \,\downarrow_6^1 \, \downarrow_{7}^3 \,\downarrow_{8}^1 \,\uparrow^2_{9} \,\downarrow^2_{10}}$}
\label{fig_paths}
\end{figure}
Examples of those walks are shown in Fig.~\ref{fig_paths}.
The ground states for the Motzkin and Fredkin Hamiltonians can be written as
follows
\bea
&&\Ket{GS}_M
=\frac{1}{\sqrt{{\cal M}^{(L)}}}\sum_{p}\Ket{{m}^{(L)}_p}\\
&&\Ket{GS}_D
=\frac{1}{\sqrt{{\cal D}^{(L)}}}\sum_p\Ket{{d}^{(L)}_p}
\eea
where the sum runs over all possible paths allowed by the length $L$ and 
colors $q$, whose numbers are given by the colored Motzkin number 
${\cal M}^{(L)}$ and the colored Catalan number ${\cal D}^{(L)}$.

\section{Combinatorics and alternative expressions for magnetization and correlation functions}
Let us define $p_n=(1-\textrm{mod}(n,2))$ such that $p_{2n+1}=0$ and $p_{2n}=1$, 
namely selects only even integer numbers, and
\be
{\cal D}^{(n)}_{hh'}=q^{\frac{n+h'-h}{2}}\left[\left(\ba{c}
n\\\frac{n+|h-h'|}{2}\ea\right)-
\left(\ba{c}
n\\\frac{n+h+h'}{2}+1\ea\right)\right]p_{n+h+h'}
\ee
where ${\cal D}^{(n)}_{hh'}$ are the number of colored Dyck-like paths 
($q$ the number of colors) between two poins with distance 
$n$ and heights $h$ and $h'$. In particular
\bea
{\cal D}^{(n)}\equiv {\cal D}^{(n)}_{00}=q^{\frac{n}{2}}C\left(\frac{n}{2}\right)\,p_n
\eea
with $C(n)=\frac{2n!}{n!(n+1)!}$ the Catalan numbers, and
\be
{\cal D}^{(n)}_{h0}=q^{\frac{n-h}{2}}\frac{h+1}{\frac{n+h}{2}+1}
\left(                               
\ba{c}                                                                         
n\\                                                                            
\frac{n+h}{2}                                                                  
\ea                                                                            
\right)p_{n+h}
\ee
and ${\cal D}^{(n)}_{0h}=q^{h}{\cal D}^{(n)}_{h0}$. Let us also define 
\be
{\cal M}^{(n)}_{h h'}=\sum_{\ell=0}^{\lf\frac{n-|h'-h|}{2}\rf}
\left(\ba{c} n\\2\ell+|h'-h|\ea\right)
{\cal D}^{(2\ell+|h'-h|)}_{hh'}
\ee
the number of colored Motzkin-like paths between two points at heights 
$h$ and $h'$. In particular
\be
{\cal M}^{(n)}\equiv {\cal M}^{(n)}_{00}=\sum_{\ell=0}^{\lfloor\frac{n}{2}\rfloor}q^\ell
\left(\ba{c} n\\
2\ell\ea\right)
C(\ell)
\ee
For $q=1$, the one-color case, ${\cal M}^{(n)}$ are the Motzkin numbers.

\begin{figure}[ht!]
\includegraphics[width=6.5cm]{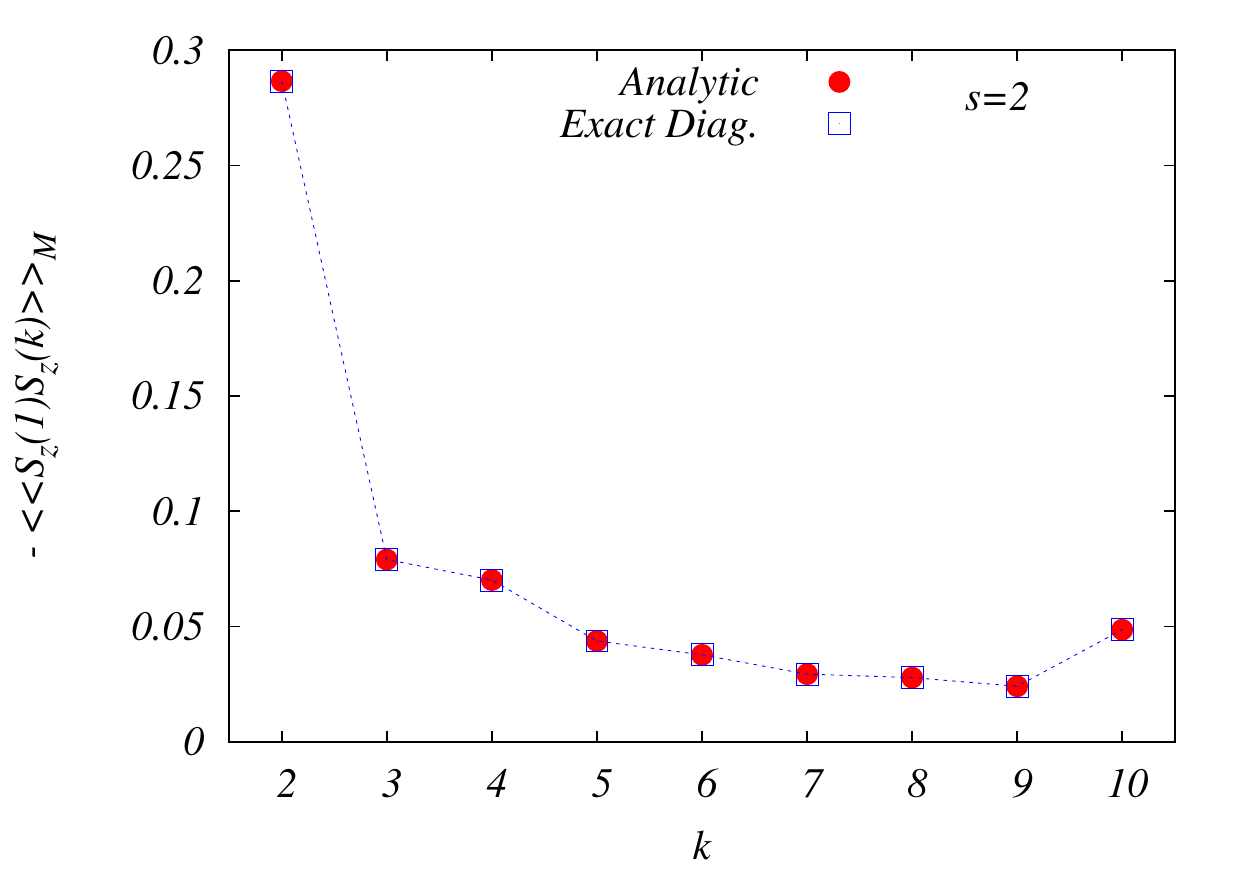}
\includegraphics[width=6.5cm]{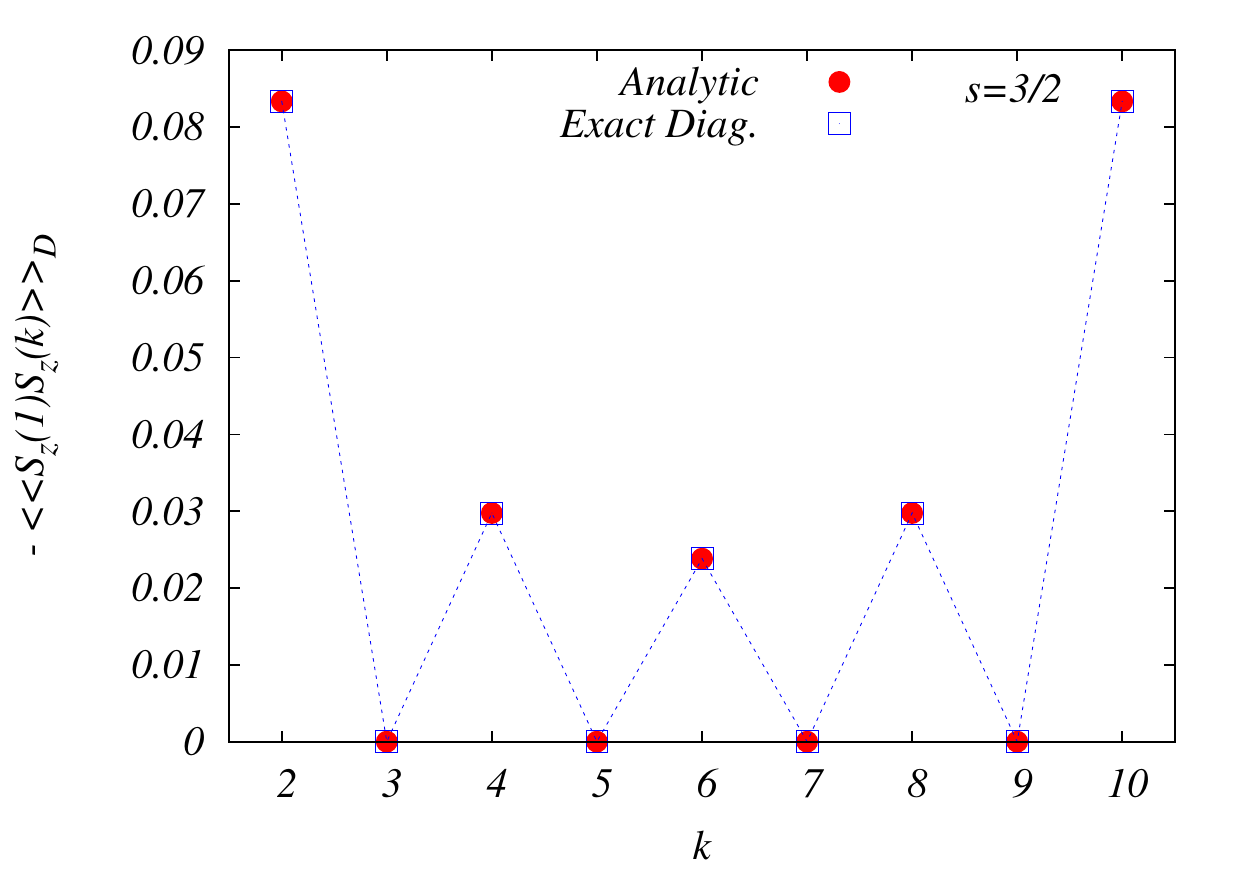}
\caption{\change{Connected correlation function $\langle S_z(1)\rangle_M
\langle S_z(k)\rangle_M-\langle S_z(1)S_z(k)\rangle_M$ for the Motzkin chain 
with $s=2$, from the analytic expressions given by Eqs.~(\ref{SzM}),~(\ref{SzSz_int2}) (or equivalently by Eqs.~(3),~(4) of the main text) and numerical exact diagonalization (upper panel), and $\langle S_z(1)\rangle_D
\langle S_z(k)\rangle_D-\langle S_z(1)S_z(k)\rangle_D$ 
for the Fredkin chain with $s=3/2$, 
from the analytic expressions given by Eqs.~(\ref{SzD}),~(\ref{SzSz_hint2}) 
(or equivalently by Eqs.~(9),~(10) 
of the main text) and numerical exact diagonalization (lower panel).
In both cases the length of the chain is $L=10$.}}
\label{fig_ned}
\end{figure}

We can now calculate the magnetization $\langle S_z(j)\rangle$ along the chains and the correlations $\langle S_z(j)S_z(k)\rangle$, whose expressions are 
given by 
Eqs.~(3-4), (9-10) of the main text. 
We can equivalently write those quantities in terms of average height or height-height correlations as follows.
For the integer spin chain ($s=q$) we have the following magnetization
\be
\langle S_z(j)\rangle_M
=\frac{(1+q)}{2{\cal M}^{(L)}}
\sum_{h} h\left({\cal M}_{0 h}^{(j)}{\cal M}_{h 0}^{(L-j)}-{\cal M}_{0 h}^{(j-1)}{\cal M}_{h 0}^{(L-j+1)}\right) 
\label{SzM}
\ee
while for half-integer spin chain ($s=\frac{2q-1}{2}$) it reads
\be
\langle S_z(j)\rangle_D
=\frac{q}{2{\cal D}^{(L)}}
\sum_{h} h\left({\cal D}_{0 h}^{(j)}{\cal D}_{h 0}^{(L-j)}-{\cal D}_{0 h}
^{(j-1)}{\cal D}_{h 0}^{(L-j+1)}\right)
\label{SzD}
\ee
The correlation function for integer and half-integer spin chains, for $k>j$, can be written as follows
\begin{widetext}
\bea
\nonumber\langle S_z(j)S_z(k)\rangle_M
=\frac{(1+q)^2}{4{\cal M}^{(L)}}
\sum_{hh'}h h'\left[{{\cal M}}^{(j)}_{0h}
{\cal M}_{h,h'}^{(k-j)}{{\cal M}}^{(L-k)}_{h'0}-{{\cal M}}^{(j)}_{0h}
{\cal M}_{h,h'}^{(k-j-1)}{{\cal M}}^{(L-k+1)}_{h'0} 
\right.\\ \nonumber \left.
-{{\cal M}}^{(j-1)}_{0h}{\cal M}_{h,h'}^{(k-j+1)}
{{\cal M}}^{(L-k)}_{h'0}
+{{\cal M}}^{(j-1)}_{0h}
{\cal M}_{h,h'}^{(k-j)}{{\cal M}}^{(L-k+1)}_{h'0}\right]\\  
-\frac{{{\cal M}}^{(k-j-1)}}{{{\cal M}}^{(L)}}\frac{(q^3-q)}{12}
\sum_{h}{{\cal M}}^{(j-1)}_{0h}{{\cal M}}^{(L-k)}_{h0}
\label{SzSz_int2}
\eea
\bea
\nonumber\langle S_z(j)S_z(k)\rangle_D
=\frac{q^2}{4{\cal D}^{(L)}}
\sum_{hh'}h h'\left[{{\cal D}}^{(j)}_{0h}
{\cal D}_{h,h'}^{(k-j)}{{\cal D}}^{(L-k)}_{h'0}-{{\cal D}}^{(j)}_{0h}
{\cal D}_{h,h'}^{(k-j-1)}{{\cal D}}^{(L-k+1)}_{h'0} 
\right.\\\nonumber \left.
-{{\cal D}}^{(j-1)}_{0h}{\cal D}_{h,h'}^{(k-j+1)}
{{\cal D}}^{(L-k)}_{h'0}+{{\cal D}}^{(j-1)}_{0h}
{\cal D}_{h,h'}^{(k-j)}{{\cal D}}^{(L-k+1)}_{h'0}\right]\\
-\frac{{{\cal D}}^{(k-j-1)}}{{{\cal D}}^{(L)}}\frac{(q^3-q)}{12}
\sum_{h}{{\cal D}}^{(j-1)}_{0h}{{\cal D}}^{(L-k)}_{h0}
\label{SzSz_hint2}
\eea
\end{widetext}
{\change {
We have compared our analytic results against DMRG and numerical exact diagonalization. Examples of such a comparison can be seen in Fig.~{\ref{fig_ned}} 
which shows perfect agreement between analytics and numerical exact diagonalization.}}

\section{Violation of the cluster decomposition}
Let us consider correlation between two sites far apart. 
For instance let us consider the site $j$ close to the left end 
of the chain and $k$ far apart, close to the other end on the right hand 
side of the chain, and send $|k-j|\rightarrow \infty$. 
We observe that the first terms in Eqs.~(4-10) of the main text 
satisfy the cluster decomposition priciple. The violation 
of the clustering is due to the second terms in those equations
\bea
&&\hspace{-0.7cm}
\nonumber 
{\langle\langle} S_z(j)S_z(k)\rangle\rangle_M
\equiv 
\Big(\langle S_z(j)S_z(k)\rangle_M
-\langle S_z(j)\rangle_M
\langle S_z(k)\rangle_M
\Big)\\
&&
\underset{|k-j|\gg 1}{\longrightarrow}
\frac{{{\cal M}}^{(k-j-1)}}{{{\cal M}}^{(L)}}\frac{(q-q^3)}{12}
\sum_{h}{{\cal M}}^{(j-1)}_{0h}{{\cal M}}^{(L-k)}_{h0}
\\
&&\hspace{-0.7cm}
\nonumber
{\langle\langle} S_z(j)S_z(k)\rangle\rangle_D
\equiv
\Big(\langle S_z(j)S_z(k)\rangle_D
-\langle S_z(j)\rangle_D
\langle S_z(k)\rangle_D
\Big)\\
&&
\underset{|k-j|\gg 1}{\longrightarrow}
\frac{{{\cal D}}^{(k-j-1)}}{{{\cal D}}^{(L)}}\frac{(q-q^3)}{12}
\sum_{h}{{\cal D}}^{(j-1)}_{0h}{{\cal D}}^{(L-k)}_{h0}
\eea
%
In particular, for $j=1$, and $1<k\le L$, for the half-integer case we get simply 
\bea
{\langle\langle} S_z(1)S_z(k)\rangle\rangle_D
&=&\frac{(q-q^3)}{12}
\frac{{{\cal D}}^{(L-k)}{{\cal D}}^{(k-2)}}{{{\cal D}}^{(L)}}\\
\nonumber
&=&\frac{(1-q^2)}{12} p_k \frac{C(\frac{L-k}{2})C(\frac{k}{2}-1)}{C(\frac{L}{2})}
\label{S1Sk}
\eea
which for $q=1$ (spin $s=1/2$) is identically equal to zero, since the spin at the first site has to be $\uparrow$, no matter the rest of the chain.
%
If we now put $k=L$ and send $L\rightarrow \infty$ we have a finite 
correlation given by Eq.~(12) in the main text 
and, more generally, one can calculate the correlators 
with $j\ge 1$ or site-symmetrical correlations as shown in 
Eqs.~(13-14) of the main text.\\
%
Analogously, for the integer spin case we get
\be
{\langle\langle} S_z(j)S_z(L)\rangle\rangle_M
\underset{L\gg j}{\longrightarrow} \frac{(q-q^3)}{12}
\frac{{{\cal M}}^{(L-j-1)}{{\cal M}}^{(j-1)}}{{{\cal M}}^{(L)}}
\ee
together with Eqs.~(5-6) of the main text.
In particular, for spin $s=q=2$ we get a finite boundary connected correlation at infinite distance, $\lim_{L\rightarrow \infty} {\langle\langle} S_z(1)S_z(L)\rangle\rangle_M
=\frac{1}{2}\lim_{L\rightarrow \infty}\frac{{\cal M}^{(L-2)}}{{\cal M}^{(L)}}\simeq -0.034$.

\section{Entanglement entropy and violation of the area law}
Let us consider a bipartition of the chain $[1,j]$, $[j+1,L]$, after Schmidt decomposition, and defining
\bea
&&P_{M}(j,h)=\frac{{\cal M}_{0 h}^{(j)}{\cal M}_{h 0}^{(L-j)}}{{\cal M}^{(L)}}\\
&&P_{D}(j,h)=\frac{{\cal D}_{0 h}^{(j)}{\cal D}_{h 0}^{(L-j)}}{{\cal D}^{(L)}}
\eea
the entanglement entropies for the two cases can be written as follows
\bea
&&{\cal S}_M
=-\sum_h P_M(j,h)
\log_2\left[q^{-h}P_M(j,h)\right]\\
&&{\cal S}_D
=-\sum_h P_D(j,h)
\log_2\left[q^{-h}P_D(j,h)\right]
\eea
Let us define
\be
{\cal S}={\cal S}^{0}+\delta {\cal S}
\ee
and one can verify that
\bea
&&{\cal S}^{0}_M
=-\sum_h P_M(j,h)
\log_2\left[P_M(j,h)\right]\\
&&{\cal S}^{0}_D
=-\sum_h P_D(j,h)
\log_2\left[P_D(j,h)\right]
\eea
fulfill the area law with standard logarithmic corrections. 
The terms which produce the violation of the area law, instead, are
\bea
&&\hspace{-0.35cm}
\delta {\cal S}_M
=\log_2(q)\,\sum_h h \, P_M(j,h)=\log_2(q)\,\langle h_j\rangle_M
\\
&&\hspace{-0.35cm}
\delta {\cal S}_D
=\log_2(q)\,\sum_h h \,P_D(j,h)=\log_2(q)\,\langle h_j\rangle_D
\eea
namely, the leading term of the entropy, which violates the area law, 
is proportional to the average height 
of the Motzkin paths (for the integer case) and the average height of 
the Dyck paths (for the half-integer case), measured right  
at the bipartition position $j$. In both cases
$\langle h_j\rangle \approx \sqrt{\frac{2j\left(L-j\right)}{L}}$.


\section{Comparison with other spin models}
\begin{figure}[ht!]
\includegraphics[width=8.0cm]{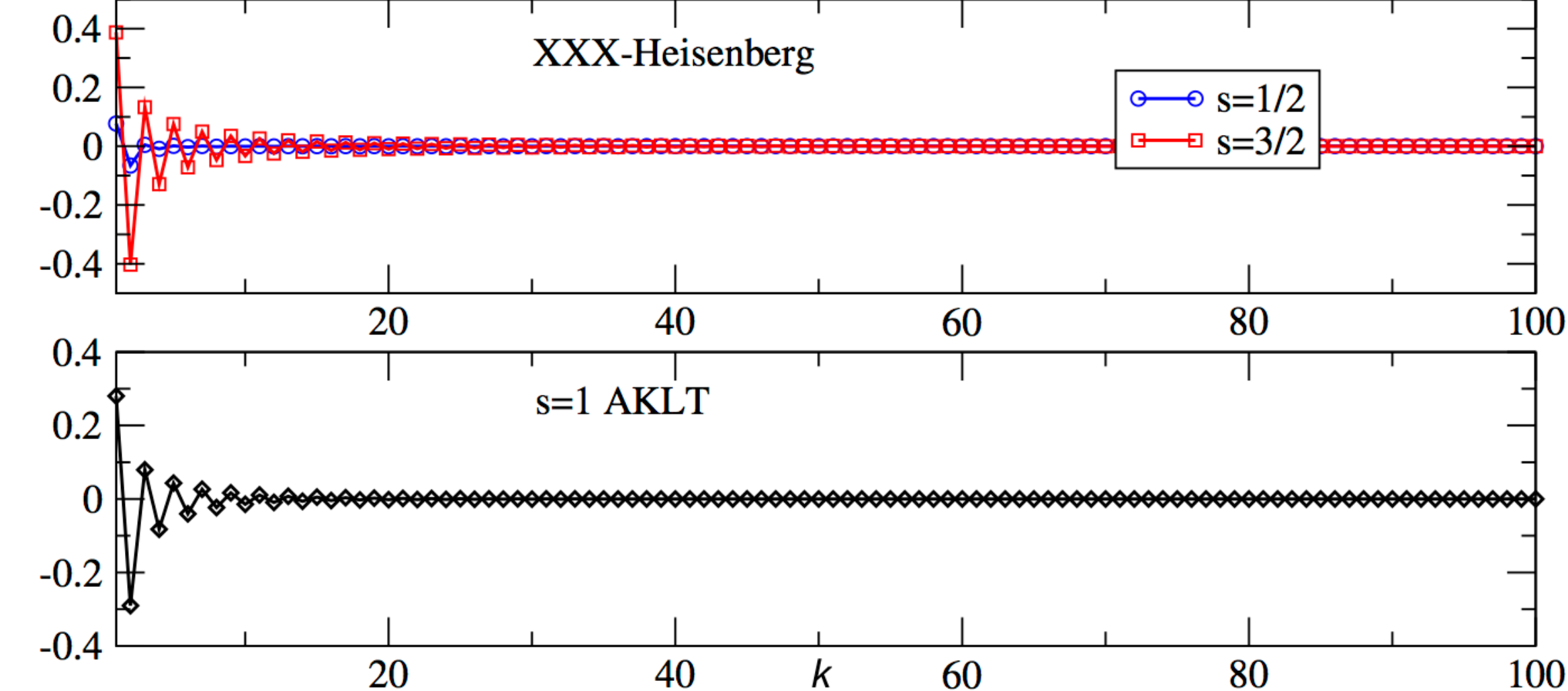}
\caption{
\change{Connected correlators $\langle\langle S_z(1)S_z(k)\rangle\rangle$ in: (upper panel) Heisenberg $XXX$ model with spins $s=1/2$ and $3/2$, 
(
lower panel) AKLT model with spin $s=1$.
}}
\label{figure_cor_sup_mat}
\end{figure}

\change{
As we stressed in the main text CDP violation is given only when the connected correlation function
\begin{equation}
\label{eqcorr}
\langle\langle S_z(j)S_z(k)\rangle\rangle=\langle S_z(j)S_z(k)\rangle-\langle S_z(j)\rangle\langle S_z(k)\rangle
\end{equation}
goes to finite value different from zero, for large $|j-k|$ distance. 
Crucially the presence of edge states \cite{aklt} 
in spin models could give rise to finite
value for large distances of the unconnected correlator $\langle S_z(j)S_z(k)\rangle$ but the CDP still holds. In order to prove that in Fig.~\ref{figure_cor_sup_mat} we calculate, by means of DMRG, 
the connected correlation function for different models.
}
\begin{figure}[ht!]
\includegraphics[width=3.6cm]{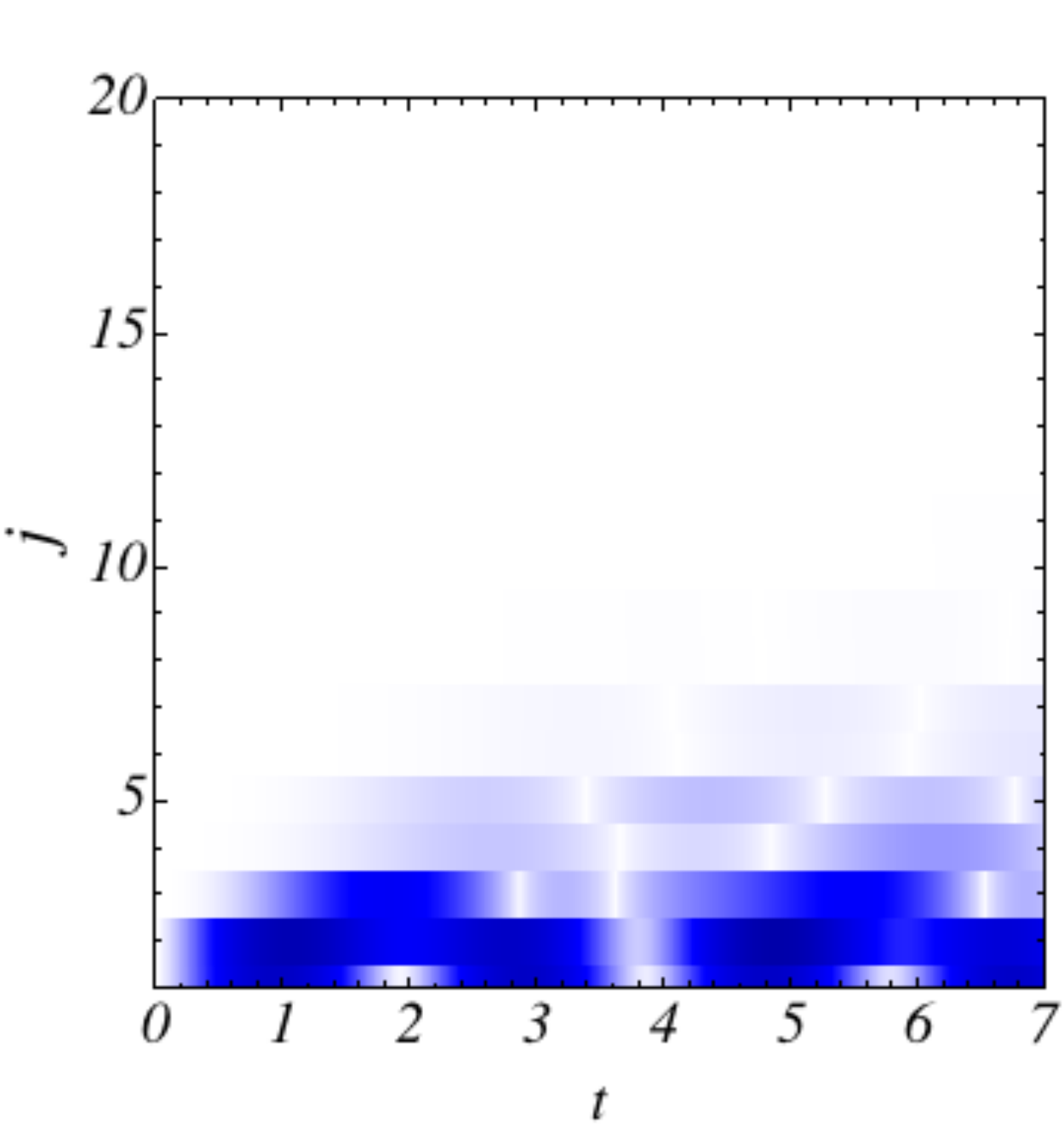}
\caption{
\change{Time evolution of $\langle S_z(j,t)\rangle-\langle S_z(j,0)\rangle$, 
for the dimerized-frustrated model~\cite{campos} (with 
$\delta=0.8$ and $\alpha=0$ following the notation in Ref.~\cite{campos}), 
after switching on a local field, $5\,S_z(j_0)$, at $j_0=0$.}}
\label{campos_fig}
\end{figure}
\change{
In the upper panel of Fig.~\ref{figure_cor_sup_mat}, Eq.~(\ref{eqcorr}) is plotted for the Heisenberg $XXX$ model both for $s=1/2$ and for $s=3/2$. 
It is clearly visible that the correlator goes to zero irrespectively of the $s$ value. 
The same behavior is observed in the lower 
panel where AKLT model for spin $s=1$ having edge states is 
treated \cite{note}. 
Finally, in Fig.~\ref{campos_fig} we plot the time evolution of the magnetization for the dimerized frustrated model diplaying long-distance entanglement \cite{campos}finding that, even if weak, a light cone-like propagation is present.  
}


\begin{thebibliography}{99}
\bibitem{note_numerics} In all our dynamical simulations the ground state is found by keeping up to 1400 DMRG states and performing 5 finite size sweeps. The dynamics is computed by using up to 1024 DMRG states and a time step $\delta t=0.01$.
\bibitem{note} Notice that both the Heisenberg XXX model and the AKLT model have degenerate ground states. Here we break the degeneracy by applying magnetic fields at the border, $H_\partial$.
\end{thebibliography}
\end{document}